\documentclass[a4paper,11pt]{JHEP3}

\usepackage{amssymb}
\usepackage{citesort}

\author{
Leszek Hadasz\footnote{Alexander von Humboldt Fellow;
    e-mail: hadasz@th.if.uj.edu.pl} \\
    Physikalisches Institut,
    Rheinische Friedrich-Wilhelms-Universit\"at,
    Nu\ss{}allee 12, 53115~Bonn, Germany
    \\
    {\rm and} \\
    M. Smoluchowski Institute of Physics,
    Jagiellonian University,
    Reymonta 4,
    30-059~Krak\'ow, Poland}

\abstract{We propose an exact form of the fusion matrix of the Neveu-Schwarz blocks
that appear in a correlation function of four super-primary fields.
Orthogonality relation satisfied by this matrix is equivalent to the bootstrap
equation for the four-point super-primary correlator in the N=1 supersymmetric Liouville
theory.}

\title{On the fusion matrix of the $N=1$ Neveu-Schwarz blocks}

\preprint{}


\begin{document}

\section{Introduction}
In the BPZ formulation of the conformal field theory \cite{Belavin:1984vu}
the basic dynamical principle is an associativity of the operator product algebra.
Its direct consequence is the bootstrap equation for the four-point correlation function \cite{Belavin:1984vu}.
Once a three point coupling constants of a given CFT are known, the bootstrap equation can
be viewed as the basic consistency condition of the theory.

The simplest CFT with a continuous spectrum, which cannot be obtained from a free field theory
in a simple way, is the Liouville theory.
Its three point coupling constants have been found independently
by Dorn and Otto \cite{Dorn:1994xn} and by Zamolodchikov and Zamolodchikov
\cite{Zamolodchikov:1995aa}. The authors of \cite{Zamolodchikov:1995aa}
also performed a numerical check of the bootstrap equation in the Liouville
theory using a recursive representation of conformal blocks developed in
\cite{Zamolodchikov:1,Zamolodchikov:2,Zamolodchikov:3}.

An analytic proof of this equation which combined  a Moore-Seiberg formalism of CFT
\cite{Moore:1988qv} with a representation theory of quantum groups
has been presented in \cite{Ponsot:1999uf,Ponsot:2000mt}.
Using the results on fusion of degenerate representation of the Virasoro algebra with
generic ones the authors of \cite{Ponsot:1999uf,Ponsot:2000mt}
derived from the consistency conditions of the Moore-Seiberg type a set of
functional equations for the fusion coefficient of the conformal blocks. These equations were then shown
to be satisfied by the Racah-Wiegner coefficients
for an appropriate continuous series of representations of U${}_q({\rm sl}(2,{\mathbb R})).$

Conformal field theory  with $N=1$ supersymmetry
\cite{Zamolodchikov:1988nm,Friedan:1984rv,Bershadsky:1985dq}
can be viewed as the simplest generalization of the ``ordinary'' CFT.
Also here the three point coupling constants of the basic interacting model ---
supersymmetric extension of the Liouville theory --- are known
\cite{Poghosian:1996dw} and some numerical checks of the
bootstrap equation in the Neveu-Schwarz sector of the theory (which employed a recursive representation
of $N=1$ NS blocks developed in \cite{Hadasz:2006qb,Belavin:2006zr})
have been performed \cite{Belavin:2007gz,Belavin:2007eq}. An analytical
proof of the consistency of the $N=1$ supersymmetric Liouville theory
is however still missing.

A goal of this paper is to make a step towards such a proof. We put forward a conjecture
on an exact form of the fusion matrix for the two NS blocks that appear in a correlation function
of four super-primary fields and check
several of its properties. As a main result of the paper one may regard identities
satisfied by the ``supersymmetric'' extensions of the basic hypergeometric functions
\cite{basic:hypergeometric}, derived in Section \ref{special:functions}. These identities
provide a technical tool which should allow to complete the proof of the consistency
of $N=1$ supersymmetric Liouville theory (perhaps along the lines mentioned in the last section).

The paper is organized as follows. In Section 2 we rewrite the bootstrap equation for
the four-point correlation functions of super-primary fields in $N=1$ Liouville theory
in the form of an orthogonality relation for the fusion matrix of $N=1$ Neveu-Schwarz blocks. In Section 3 we
construct --- guided by an analogy with the form of the fusion matrix for ``ordinary'' Liouville blocks\footnote{
This analogy extends in fact to other objects is supersymmetric and ``ordinary'' Liouville theory,
including reflection amplitudes, boundary two point functions and ``bulk'' one point functions
in the disc and in the Lobachevsky plane geometries, see for instance a review article
\cite{Nakayama:2004vk}.} --- a conjectured fusion matrix for the basic NS blocks. In Section 4
we prove some of its most important properties: orthogonality, symmetry properties
and calculate its form in the case which corresponds to a degenerate representation
of the NS algebra. Section 5 contains proofs of several identities satisfied
by a ``supersymmetric extensions'' of the basic hypergeometric functions: integral analogs
of the Ramanujan summation formula, Heine and Euler-Heine transformations and
analog of Saalsch\"utz summation formulae. We end up with some discussion of
possible future applications of the results.

\section{Bootstrap in the $N=1$ supersymmetric Liouville theory}
\label{Bootstrap}
Let
\begin{eqnarray*}
C(\alpha_3,\alpha_2,\alpha_1)
& = &
C_0(\alpha)\,
\frac{\Upsilon_{\rm NS}(2\alpha_3)\Upsilon_{\rm NS}(2\alpha_2)\Upsilon_{\rm NS}(2\alpha_1)}
{\Upsilon_{\rm NS}(\alpha-Q)\Upsilon_{\rm NS}(\alpha_{1+2-3})\Upsilon_{\rm NS}(\alpha_{2+3-1})\Upsilon_{\rm NS}(\alpha_{3+1-2})},
\\[10pt]
\widetilde C(\alpha_3,\alpha_2,\alpha_1)
& = &
i\,C_0(\alpha)\,
\frac{\Upsilon_{\rm NS}(2\alpha_3)\Upsilon_{\rm NS}(2\alpha_2)\Upsilon_{\rm NS}(2\alpha_1)}
{\Upsilon_{\rm R}(\alpha-Q)\Upsilon_{\rm R}(\alpha_{1+2-3})\Upsilon_{\rm R}(\alpha_{2+3-1})\Upsilon_{\rm R}(\alpha_{3+1-2})},
\end{eqnarray*}
with
\[
C_0(\alpha) =
\left(\pi\mu\gamma\left(\frac{bQ}{2}\right)b^{1-b^2}\right)^{\frac{Q-\alpha}{b}}\Upsilon'_{\rm NS}(0),
\]
denote the two independent structure constants in the Neveu-Schwarz sector of the $N=1$ supersymmetric
Liouville theory\footnote
{The constant $\tilde C$ differs from a corresponding constant in \cite{Belavin:2007gz} by a factor of $\frac12,$
and consequently an odd NS block differs by a factor of 2 from the conventions of \cite{Belavin:2007gz}.}.
Here $\mu$ is a two-dimensional cosmological constant, $b$ denotes a Liouville coupling constant,
$Q = b + b^{-1}$ is the background charge related to the central charge of the Neveu-Schwarz algebra as
\[
c = \frac32 + 3Q^2,
\]
$\alpha \equiv \alpha_1 + \alpha_2+\alpha_3,\ \alpha_{1+2-3} \equiv \alpha_1 +\alpha_2 -\alpha_3,$ etc.\
and the special functions involved ($\Upsilon_{\rm NS,R}(x), \Gamma_{\rm NS,R}(x)$ below etc.)
 are defined in Section \ref{special:functions}.

It is convenient to combine $C$ and $\tilde C$  in the matrix notation
\[
{\sf C}(\alpha_3,\alpha_2,\alpha_1)
\;\ = \;\
\left(
\begin{array}{cc}
C(\alpha_3,\alpha_2,\alpha_1) & 0
\\[4pt]
0 & \widetilde C(\alpha_s,\alpha_2,\alpha_1)
\end{array}
\right).
\]
Let us define:
\[
\vec{\cal F}_{\!\alpha_s}\!\left[^{\alpha_3\: \alpha_2}_{\alpha_4\: \alpha_1}\right]\!(z)
\;\ = \;\
\left(
{\cal F}_{\!\alpha_s}^{\rm e}\!\left[^{\alpha_3\: \alpha_2}_{\alpha_4\: \alpha_1}\right]\!(z),
{\cal F}_{\!\alpha_s}^{\rm o}\!\left[^{\alpha_3\: \alpha_2}_{\alpha_4\: \alpha_1}\right]\!(z)
\right),
\]
where ${\cal F}_{\!\alpha_s}^{\rm e}$  and ${\cal F}_{\!\alpha_s}^{\rm o}$ denote an even and an odd $N=1$ Neveu-Schwarz block
 \cite{Belavin:2007gz,Hadasz:2006qb}.

Four-point correlation function of super-primary NS fields in the $N=1$ supersymmetric Liouville theory,
\[
G_4(z,\bar z)
=
\Big\langle
V_{\alpha_4}(\infty,\infty)
V_{\alpha_3}(1,1)
V_{\alpha_2}(z,\bar z)
V_{\alpha_1}(0,0)
\Big\rangle,
\]
can be written either in the ``$s-$channel'' representation:
\begin{eqnarray*}
G_4(z,\bar z)
& = &
\hskip -10pt
\int\limits_{\frac{Q}{2} + i{\mathbb R}_+}
\hskip -10pt \frac{d\alpha_s}{i}
\left[
C(\alpha_4,\alpha_3,\alpha_s)C(\bar\alpha_s,\alpha_2,\alpha_1)
\left|{\cal F}_{\!\alpha_s}^{\rm e}\!\left[^{\alpha_3\: \alpha_2}_{\alpha_4\: \alpha_1}\right]\!(z)\right|^2
\right.
\\
&&
\hskip .7cm
\left.
-\
\widetilde C(\alpha_4,\alpha_3,\alpha_s)\widetilde C(\bar\alpha_s,\alpha_2,\alpha_1)
\left|{\cal F}_{\!\alpha_s}^{\rm o}\!\left[^{\alpha_3\: \alpha_2}_{\alpha_4\: \alpha_1}\right]\!(z)\right|^2
\right]
\\[10pt]
& = &
\hskip -10pt
\int\limits_{\frac{Q}{2} + i{\mathbb R}_+}
\hskip -10pt \frac{d\alpha_s}{i}\
\vec{\cal F}_{\!\alpha_s}\!\left[^{\alpha_3\: \alpha_2}_{\alpha_4\: \alpha_1}\right]\!(z)\,
{\sf C}(\alpha_4,\alpha_3,\alpha_s) \cdot \sigma_3 \cdot {\sf C}(\bar\alpha_s,\alpha_2,\alpha_1)
\left(\vec{\cal F}_{\!\alpha_s}\!\left[^{\alpha_3\: \alpha_2}_{\alpha_4\: \alpha_1}\right]\!(z)\right)^{\dag},
\end{eqnarray*}
(here and in what follows we use a convenient notation $\bar\alpha = Q-\alpha;$ notice that for $\alpha \in \frac{Q}{2} + i{\mathbb R}$
it is indeed the complex conjugation od $\alpha$) or in the ``$t-$channel'' representation:
\begin{eqnarray*}
G_4(z,\bar z)
& = &
\hskip -10pt
\int\limits_{\frac{Q}{2} + i{\mathbb R}_+}
\hskip -10pt \frac{d\alpha_t}{i}\
\vec{\cal F}_{\!\alpha_t}\!\left[^{\alpha_1\: \alpha_2}_{\alpha_4\: \alpha_3}\right]\!(1-z)\,
{\sf C}(\alpha_4,\alpha_t,\alpha_1) \cdot \sigma_3 \cdot {\sf C}(\bar\alpha_t,\alpha_3,\alpha_2)
\left(\vec{\cal F}_{\!\alpha_t}\!\left[^{\alpha_1\: \alpha_2}_{\alpha_4\: \alpha_3}\right]\!(1-z)\right)^{\dag}.
\end{eqnarray*}
Coincidence of these two representations constitute the bootstrap equation
for the super-primary fields in the supersymmetric Liouville field theory (SLFT).

Defining the SLFT fusion matrix
\(
{\sf F}_{\alpha_s\alpha_t}\!\left[^{\alpha_3\: \alpha_2}_{\alpha_4\: \alpha_1}\right]
\)
through the equation:
\begin{equation}
\label{Fusion:definition}
\vec {\cal F}_{\alpha_s}\!\left[^{\alpha_3\: \alpha_2}_{\alpha_4\: \alpha_1}\right]\!(z)
\;\ = \;\
\hskip -10pt
\int\limits_{\frac{Q}{2} + i{\mathbb R}_+}
\hskip -10pt \frac{d\alpha_t}{i}\
\vec {\cal F}_{\alpha_t}\!\left[^{\alpha_1\ \alpha_2}_{\alpha_4\ \alpha_3}\right]\!(1-z)
\,
{\sf F}_{\alpha_s\alpha_t}\!\left[^{\alpha_3\: \alpha_2}_{\alpha_4\: \alpha_1}\right],
\end{equation}
we can rewrite the bootstrap equation in the form of an orthogonality relation:
\begin{eqnarray}
\label{orthogonality:F}
\nonumber
&&
\hskip -2cm
\int\limits_{\frac{Q}{2} + i{\mathbb R}_+}
\hskip -10pt \frac{d\alpha_s}{i}\
{\sf F}_{\alpha_s\alpha_t}\!\left[^{\alpha_3\: \alpha_2}_{\alpha_4\: \alpha_1}\right]
\cdot
{\sf C}(\alpha_4,\alpha_3,\alpha_s) \cdot \sigma_3 \cdot {\sf C}(\bar\alpha_s,\alpha_2,\alpha_1)
\cdot
\left({\sf F}_{\alpha_s\alpha'_t}\!\left[^{\alpha_3\: \alpha_2}_{\alpha_4\: \alpha_1}\right]\right)^\dag
\\[-6pt]
&&
\hskip 1.1cm = \;\
{\sf C}(\alpha_4,\alpha_t,\alpha_1) \cdot \sigma_3 \cdot {\sf C}(\bar\alpha_t,\alpha_3,\alpha_2)\
i\delta(\alpha_t-\alpha'_t),
\end{eqnarray}
where
\[
\sigma_3
\equiv
\left(
\begin{array}{rr}
1 & 0 \\[-4pt] 0 & -1
\end{array}
\right).
\]

Define now
\begin{eqnarray}
\nonumber
{\cal N}_{\rm NS}(\alpha_3,\alpha_2,\alpha_1)
& = &
\frac{\Gamma_{\rm NS}(2Q-2\alpha_3)\Gamma_{\rm NS}(2\alpha_2)\Gamma_{\rm NS}(2\alpha_1)}
{\Gamma_{\rm NS}(2Q-\alpha_{3+2+1})\Gamma_{\rm NS}(\alpha_{3+2-1})
\Gamma_{\rm NS}(\alpha_{2+1-3})\Gamma_{\rm NS}(\alpha_{1+3-2}) },
\\
\nonumber
{\cal N}_{\rm R}(\alpha_3,\alpha_2,\alpha_1)
& = &
\frac{\Gamma_{\rm NS}(2Q-2\alpha_3)\Gamma_{\rm NS}(2\alpha_2)\Gamma_{\rm NS}(2\alpha_1)}
{\Gamma_{\rm R}(2Q-\alpha_{3+2+1})\Gamma_{\rm R}(\alpha_{3+2-1})
\Gamma_{\rm R}(\alpha_{2+1-3})\Gamma_{\rm R}(\alpha_{1+3-2}) },
\end{eqnarray}
\begin{eqnarray}
\label{normalization:N}
{\sf N}_{\alpha_s}\!\left[^{\alpha_3\: \alpha_2}_{\alpha_4\: \alpha_1}\right]
& = &
\left(
\begin{array}{cc}
{\cal N}_{\rm NS}(\alpha_4,\alpha_3,\alpha_s){\cal N}_{\rm NS}(\alpha_s,\alpha_2,\alpha_1) &  \hskip -20pt 0
\\[10pt]
0 & \hskip -20pt {\cal N}_{\rm R}(\alpha_4,\alpha_3,\alpha_s){\cal N}_{\rm R}(\alpha_s,\alpha_2,\alpha_1)
\end{array}
\right),
\end{eqnarray}
and
\begin{eqnarray}
\label{FusionG:definition}
{\sf G}_{\alpha_s\alpha_t}\!\left[^{\alpha_3\: \alpha_2}_{\alpha_4\: \alpha_1}\right]
& = &
{\sf N}_{\alpha_t}\!\left[^{\alpha_1\ \alpha_2}_{\alpha_4\ \alpha_3}\right]
\cdot
{\sf F}_{\alpha_s\alpha_t}\!\left[^{\alpha_3\: \alpha_2}_{\alpha_4\: \alpha_1}\right]
\cdot
\left({\sf N}_{\alpha_s}\!\left[^{\alpha_3\: \alpha_2}_{\alpha_4\: \alpha_1}\right]\right)^{-1}.
\end{eqnarray}
These definitions are motivated by an identity:
\begin{eqnarray}
\nonumber
&&
\hskip -2cm
{\sf N}_{s}\!\left[^{\alpha_3\: \alpha_2}_{\alpha_4\: \alpha_1}\right]
{\sf C}(\alpha_4,\alpha_3,\alpha_s)\cdot\sigma_3\cdot {\sf C}(\bar\alpha_s,\alpha_2,\alpha_1)
\left({\sf N}_{s}\!\left[^{\alpha_3\: \alpha_2}_{\alpha_4\: \alpha_1}\right]\right)^{\dag}
\\[6pt]
\label{C:through:N}
& = &
\left(\pi\mu\gamma\left(\frac{bQ}{2}\right)b^{1-b^2}\right)^{\frac{Q-\alpha_{1+2+3+4}}{b}}
\left(\Upsilon'_{\rm NS}(0)\right)^2
\left|S_{\rm NS}(2\alpha_s)\right|^2
\sigma_0,
\end{eqnarray}
where
\[
\sigma_0
\equiv
\left(
\begin{array}{cc}
1 & 0 \\[-4pt] 0 & 1
\end{array}
\right).
\]
In view of (\ref{C:through:N})  we can rewrite the equation (\ref{orthogonality:F}) as a simple orthogonality
relation for the matrix $\sf G:$
\begin{equation}
\label{orthogonality:G}
\int\limits_{\frac{Q}{2} + i{\mathbb R}_+}
\hskip -10pt \frac{d\alpha_s}{i}\
\left|S_{\rm NS}(2\alpha_s)\right|^2\,
{\sf G}_{\alpha_s\alpha_t}\!\left[^{\alpha_3\: \alpha_2}_{\alpha_4\: \alpha_1}\right]
\cdot
\left({\sf G}_{\alpha_s\alpha'_t}\!\left[^{\alpha_3\: \alpha_2}_{\alpha_4\: \alpha_1}\right]\right)^\dag
=
\left|S_{\rm NS}(2\alpha_t)\right|^2
\sigma_0\
i\delta(\alpha_t-\alpha'_t).
\end{equation}

The fusion matrix ${\sf F}_{\alpha_s\alpha_t}\!\left[^{\alpha_3\: \alpha_2}_{\alpha_4\: \alpha_1}\right]$
is expected \cite{Moore:1988qv} to be invariant with respect to the  separate conjugations $\alpha_i \to Q-\alpha_i$
of all six of its arguments
(and thus to depend only on the conformal weights $\Delta_i = \frac12 \alpha_i(Q-\alpha_i)$) and should not change
under exchange of its rows and columns,
\[
{\sf F}_{\alpha_s\alpha_t}\!\left[^{\alpha_3\: \alpha_2}_{\alpha_4\: \alpha_1}\right]
\; = \;
{\sf F}_{\alpha_s\alpha_t}\!\left[^{\alpha_4\: \alpha_1}_{\alpha_3\: \alpha_2}\right]
\; = \;
{\sf F}_{\alpha_s\alpha_t}\!\left[^{\alpha_2\: \alpha_3}_{\alpha_1\: \alpha_4}\right].
\]

\section{Explicit form of the fusion matrix}
In analogy with \cite{Ponsot:1999uf,Ponsot:2000mt} we shall define a ``supersymmetric'' deformed hypergeometric function:
\begin{eqnarray}
\label{deformedF:definition}
{\mathbb F}(\alpha,\beta;\gamma;z)
& = &
\left(
\begin{array}{rr}
F_{\rm NS}^{(+)}(\alpha,\beta;\gamma;z) &\; F_{\rm R}^{(-)}(\alpha,\beta;\gamma;z)
\\[10pt]
F_{\rm NS}^{(-)}(\alpha,\beta;\gamma;z) &\; F _{\rm R}^{(+)}(\alpha,\beta;\gamma;z)
\end{array}
\right),
\end{eqnarray}
where\footnote{
The contour of integration is located to the right
of the poles of the integrand coming from the poles of the functions in the numerator
and to the left from the poles of the integrand due to zeroes of the functions in the denominator;
see definitions in Section \ref{special:functions}.}:
\begin{eqnarray*}
F_{\rm NS}^{(\pm)}(\alpha,\beta;\gamma;z)
& = &
\frac{S_{\rm NS}(\gamma)}{S_{\rm NS}(\alpha)S_{\rm NS}(\beta)}
\\
& \times &
\int\limits_{-i\infty}^{i\infty}\!\frac{d\tau}{i}\
{\rm e}^{i\pi\tau z}
\left[
\frac{S_{\rm NS}(\tau + \alpha) S_{\rm NS}(\tau + \beta)}{S_{\rm NS}(\tau + \gamma) S_{\rm NS}(\tau + Q)}
\pm
\frac{S_{\rm R}(\tau + \alpha) S_{\rm R}(\tau + \beta)}{S_{\rm R}(\tau + \gamma) S_{\rm R}(\tau + Q)}
\right],
\\[6pt]
F_{\rm R}^{(\pm)}(\alpha,\beta;\gamma;z)
& = &
\frac{S_{\rm NS}(\gamma)}{S_{\rm R}(\alpha)S_{\rm R}(\beta)}
\\
& \times &
\int\limits_{-i\infty}^{i\infty}\!\frac{d\tau}{i}\
{\rm e}^{i\pi\tau z}
\left[
\frac{S_{\rm NS}(\tau + \alpha) S_{\rm NS}(\tau + \beta)}{S_{\rm R}(\tau + \gamma) S_{\rm R}(\tau + Q)}
\pm
\frac{S_{\rm R}(\tau + \alpha) S_{\rm R}(\tau + \beta)}{S_{\rm NS}(\tau + \gamma) S_{\rm NS}(\tau + Q)}
\right].
\end{eqnarray*}
It satisfies an important relation
\begin{equation}
\label{F:reflection}
{\mathbb F}(\alpha,\beta;\gamma;z)
=
{\rm e}^{\frac{i\pi}{2}(\gamma-\alpha-\beta)z}
\left(
\begin{array}{cc}
\frac{S_{\rm NS}(z+\frac{Q+\gamma-\alpha-\beta}{2})}{S_{\rm NS}(z+\frac{Q-\gamma+\alpha+\beta}{2})} & 0
\\
0 & \frac{S_{\rm R}(z+\frac{Q+\gamma-\alpha-\beta}{2})}{S_{\rm R}(z+\frac{Q-\gamma+\alpha+\beta}{2})}
\end{array}
\right)
{\mathbb F}(\gamma - \alpha,\gamma - \beta;\gamma;z),
\end{equation}
which arises by expressing (\ref{Euler:Heine}) through the functions $S_{\rm NS}$ and $S_{\rm R}.$

Let us further define
\begin{eqnarray}
\nonumber
\Theta_s(x|\alpha_s)
& = &
\frac{1}{4\sqrt 2}\,
{\rm e}^{\pi x(\frac{Q}{2} +\alpha_s -\alpha_1-\alpha_2)}
\left(
\begin{array}{rr}
1 & 1
\\[-4pt]
1 & -1
\end{array}
\right)
{\mathbb F}(\alpha_s+\alpha_1-\alpha_2,\alpha_s+\alpha_3-\bar\alpha_4;2\alpha_s;-ix),
\\[-6pt]
\label{Theta:throuh:F}
\\[-6pt]
\nonumber
\Theta_t(x|\alpha_t)
& = &
\frac{1}{4\sqrt 2}\,
{\rm e}^{-\pi x(\alpha_t +\alpha_1-\bar\alpha_4-\frac{Q}{2})}
\left(
\begin{array}{rr}
1 & 1
\\[-4pt]
1 & -1
\end{array}
\right)
{\mathbb F}(\alpha_t+\alpha_1-\bar\alpha_4,\alpha_t+\alpha_3-\alpha_2;2\alpha_t;ix).
\end{eqnarray}
Explicitly,
\[
\Theta_s(x|\alpha_s)
\; = \;
\frac1{2\sqrt 2}\, S_{\rm NS}(2\alpha_s)
\left(
\begin{array}{rr}
\theta^{s}_{\rm\scriptscriptstyle NN}(x|\alpha_s) & \theta^{s}_{\rm\scriptscriptstyle NR}(x|\alpha_s)
\\[10pt]
\theta^{s}_{\rm\scriptscriptstyle RR}(x|\alpha_s) & -\,\theta^{s}_{\rm\scriptscriptstyle RN}(x|\alpha_s)
\end{array}
\right)\cdot {\mathbf U}_{\!s}(\alpha_s)
\]
with
\[
{\mathbf U}_{\!s}(\alpha_s)
\;\ = \;\
\left(
\begin{array}{cc}
S_{\rm NS}(\alpha_s+\alpha_1-\alpha_2)S_{\rm NS}(\alpha_s+\alpha_3-\bar\alpha_4) &
\hskip -5mm 0
\\[10pt]
0 & \hskip -5mm
S_{\rm R}(\alpha_s+\alpha_1-\alpha_2)S_{\rm R}(\alpha_s+\alpha_3-\bar\alpha_4)
\end{array}
\right)^{-1}
\hskip -3mm
\]
and
\begin{eqnarray}
\nonumber
\theta^{s}_{\rm\scriptscriptstyle NN}(x|\alpha_s)
& = &
\int\limits_{-i\infty}^{i\infty}\! \frac{d\tau}{i}\
{\rm e}^{\pi x(\frac{Q}{2}+\tau+\alpha_s-\alpha_1-\alpha_2)}\
\frac{S_{\rm NS}(\tau + \alpha_s+\alpha_1-\alpha_2)S_{\rm NS}(\tau + \alpha_s+\alpha_3-\bar\alpha_4)}
{S_{\rm NS}(\tau + 2\alpha_s)S_{\rm NS}(\tau + Q)},
\\
\nonumber
\theta^{s}_{\rm\scriptscriptstyle NR}(x|\alpha_s)
& = &
\int\limits_{-i\infty}^{i\infty}\! \frac{d\tau}{i}\
{\rm e}^{\pi x(\frac{Q}{2}+\tau+\alpha_s-\alpha_1-\alpha_2)}\
\frac{S_{\rm NS}(\tau + \alpha_s+\alpha_1-\alpha_2)S_{\rm NS}(\tau + \alpha_s+\alpha_3-\bar\alpha_4)}
{S_{\rm R}(\tau + 2\alpha_s)S_{\rm R}(\tau + Q)},
\\[-6pt]
\label{small:thetas}
\\[-6pt]
\nonumber
\theta^{s}_{\rm\scriptscriptstyle RN}(x|\alpha_s)
& = &
\int\limits_{-i\infty}^{i\infty}\! \frac{d\tau}{i}\
{\rm e}^{\pi x(\frac{Q}{2}+\tau+\alpha_s-\alpha_1-\alpha_2)}\
\frac{S_{\rm R}(\tau + \alpha_s+\alpha_1-\alpha_2)S_{\rm R}(\tau + \alpha_s+\alpha_3-\bar\alpha_4)}
{S_{\rm NS}(\tau + 2\alpha_s)S_{\rm NS}(\tau + Q)},
\\
\nonumber
\theta^{s}_{\rm\scriptscriptstyle RR}(x|\alpha_s)
& = &
\int\limits_{-i\infty}^{i\infty}\! \frac{d\tau}{i}\
{\rm e}^{\pi x(\frac{Q}{2}+\tau+\alpha_s-\alpha_1-\alpha_2)}\
\frac{S_{\rm R}(\tau + \alpha_s+\alpha_1-\alpha_2)S_{\rm R}(\tau + \alpha_s+\alpha_3-\bar\alpha_4)}
{S_{\rm R}(\tau + 2\alpha_s)S_{\rm R}(\tau + Q)}.
\end{eqnarray}
Notice that ${\mathbf U}_{\!s}(\alpha_s)$ is $x-$independent  and (for $\alpha_s,\alpha_i \in \frac{Q}{2} + i{\mathbb R}$) unitary,
\[
{\mathbf U}^\dag_{\!s}(\alpha_s) {\mathbf U}_{\!s}(\alpha_s) \;\ = \;\ {\mathbf U}_{\!s}(\alpha_s){\mathbf U}^{\dag}_{\!s}(\alpha_s)
=
\sigma_0.
\]
Finally, define (unitary) normalization factors:
\begin{eqnarray}
\nonumber
M_\flat^s\left[^{\alpha_3\:\alpha_2}_{\alpha_4\:\alpha_1}\right]
& = &
\frac
{S_\flat(\alpha_s + \alpha_3-\alpha_4)S_\flat(\alpha_s-\alpha_1+\alpha_2)S_\flat(\alpha_s+\alpha_1-\alpha_2)}
{S_\flat(\alpha_s-\alpha_1+\bar\alpha_2)},
\hskip 1cm
\flat = {\rm NS,\, R},
\\[10pt]
\label{M:definition}
M_\flat^t\left[^{\alpha_3\:\alpha_2}_{\alpha_4\:\alpha_1}\right]
& = &
\frac
{S_\flat(\alpha_t + \alpha_1-\alpha_4)S_\flat(\alpha_t-\alpha_3+\alpha_2)S_\flat(\alpha_t+\alpha_3-\alpha_2)}
{S_\flat(\alpha_t-\alpha_3+\bar\alpha_2)},
\\[10pt]
\nonumber
{\mathbf M}^{\natural}_{\alpha_\natural}\!\left[^{\alpha_3\:\alpha_2}_{\alpha_4\:\alpha_1}\right]
& = &
\left(
\begin{array}{cc}
M^{\natural}_{\rm NS}\left[^{\alpha_3\:\alpha_2}_{\alpha_4\:\alpha_1}\right] & 0
\\[6pt]
0 & M^{\natural}_{\rm R}\left[^{\alpha_3\:\alpha_2}_{\alpha_4\:\alpha_1}\right]
\end{array}
\right),
\hskip 5cm
\natural = s,\,t.
\end{eqnarray}
Let
\begin{equation}
\label{G:definition}
{\mathbf G}_{\alpha_s\alpha_t}\!\left[^{\alpha_3\: \alpha_2}_{\alpha_4\: \alpha_1}\right]
=
\frac{S_{\rm NS}(2\alpha_t)}{S_{\rm NS}(2\alpha_s)}\,
\left({\mathbf M}^{t}_{\alpha_t}\!\left[^{\alpha_3\:\alpha_2}_{\alpha_4\:\alpha_1}\right]\right)^\dag
\left(
\int\limits_{\mathbb R}\!dx\
\Theta^\dag_t(x|\alpha_t)\Theta_s(x|\alpha_s)
\right)
{\mathbf M}^{s}_{\alpha_s}\!\left[^{\alpha_3\:\alpha_2}_{\alpha_4\:\alpha_1}\right],
\end{equation}
and
\begin{equation}
\label{F:definition}
{\mathbf F}_{\alpha_s\alpha_t}\!\left[^{\alpha_3\: \alpha_2}_{\alpha_4\: \alpha_1}\right]
=
\left({\sf N}_{\alpha_t}\!\left[^{\alpha_1\: \alpha_2}_{\alpha_4\: \alpha_3}\right]\right)^{-1}\!\!
\cdot
{\mathbf G}_{\alpha_s\alpha_t}\!\left[^{\alpha_3\: \alpha_2}_{\alpha_4\: \alpha_1}\right]
\cdot
{\sf N}_{\alpha_s}\!\left[^{\alpha_3\: \alpha_2}_{\alpha_4\: \alpha_1}\right],
\end{equation}
where the normalization factors
\(
{\sf N}_{s}\!\left[^{\alpha_3\: \alpha_2}_{\alpha_4\: \alpha_1}\right]
\)
are defined in Eq.\ (\ref{normalization:N}).

We expect that an equality:
\begin{equation}
\label{conjecture}
{\sf F}_{\alpha_s\alpha_t}\!\left[^{\alpha_3\: \alpha_2}_{\alpha_4\: \alpha_1}\right]
=
{\mathbf F}_{\alpha_s\alpha_t}\!\left[^{\alpha_3\: \alpha_2}_{\alpha_4\: \alpha_1}\right],
\end{equation}
where
\(
{\sf F}_{\alpha_s\alpha_t}\!\left[^{\alpha_3\: \alpha_2}_{\alpha_4\: \alpha_1}\right]
\)
is the fusion matrix defined in Eq.\ (\ref{Fusion:definition}), holds. In the next section we shall present some arguments
in favor of (\ref{conjecture}).

\section{Properties of the matrix $\mathbf F$}
\subsection{Orthogonality}
A short calculation allows to check  an identity:
\begin{eqnarray*}
&&
\hskip -15mm
\int\limits_{\mathbb R}\!dx\
\Theta^\dag_s(x|\alpha_s) \Theta_s(x|\alpha'_s)
= \frac14  \overline{S_{\rm NS}(2\alpha_s)} S_{\rm NS}(2\alpha_s')
\\
& \times &
{\mathbf U}^\dag_s(\alpha_s)
\int\limits_{-i\infty}^{i\infty}\! \frac{d\tau}{i}\
\left(
\begin{array}{rr}
\left\langle \tau\left|^{\rm\scriptscriptstyle N}_{\rm\scriptscriptstyle N}\right|\xi_s\right\rangle
&
\left\langle \tau\left|^{\rm\scriptscriptstyle R}_{\rm\scriptscriptstyle R}\right|\xi_s\right\rangle
\\[10pt]
\left\langle \tau\left|^{\rm\scriptscriptstyle R}_{\rm\scriptscriptstyle R}\right|\xi_s\right\rangle
&
-\left\langle \tau\left|^{\rm\scriptscriptstyle N}_{\rm\scriptscriptstyle N}\right|\xi_s\right\rangle
\end{array}
\right)^{\dag}
\left(
\begin{array}{rr}
\left\langle \tau\left|^{\rm\scriptscriptstyle N}_{\rm\scriptscriptstyle N}\right|\xi_s'\right\rangle
&
\left\langle \tau\left|^{\rm\scriptscriptstyle R}_{\rm\scriptscriptstyle R}\right|\xi_s'\right\rangle
\\[10pt]
\left\langle \tau\left|^{\rm\scriptscriptstyle R}_{\rm\scriptscriptstyle R}\right|\xi_s'\right\rangle
&
-\left\langle \tau\left|^{\rm\scriptscriptstyle N}_{\rm\scriptscriptstyle N}\right|\xi_s'\right\rangle
\end{array}
\right)
{\mathbf U}(\alpha_s'),
\end{eqnarray*}
where $\alpha_s = \frac{Q}{2} + \xi_s,\ \alpha'_s = \frac{Q}{2} + \xi'_s,\ \xi_s, \xi'_s \in i{\mathbb R}_{+}$
and the symbols
\(
\left\langle \tau\left|^{\rm\scriptscriptstyle N}_{\rm\scriptscriptstyle N}\right|\xi_s\right\rangle
\)
etc.\ are defined in Section 5.3.
Using (\ref{orthogonality:1}), (\ref{orthogonality:2}) and relations:
\[
\left|S_{\rm NS}(2\alpha_s)\right|^2
=
S_{\rm NS}(Q-2\xi_s)S_{\rm NS}(Q+2\xi_s)
=
\frac{S_{\rm NS}(2\xi_s+Q)}{S_{\rm NS}(2\xi_s)},
\]
we thus get
\begin{equation}
\label{Theta:s:orthogonality}
\int\limits_{\mathbb R}\!dx\
\Theta^\dag_s(x|\alpha_s) \Theta_s(x|\alpha'_s)
=
\sigma_0\ i\delta(\alpha_s-\alpha_s').
\end{equation}
Furthermore, since
\begin{eqnarray*}
\theta_{\rm\scriptscriptstyle NN}(x|\alpha_s)
\overline{\theta_{\rm\scriptscriptstyle NN}(y|\alpha_s)}
& = &
\int\hskip -15pt\int\limits_{-i\infty\;\;\;\;\; }^{i\infty\;}\!\!\! \frac{d\tau}{i}\frac{d\lambda}{i}\
{\rm e}^{\pi x(Q+\tau-\alpha_1-\alpha_2) + \pi y(-\lambda+\alpha_1+\alpha_2-Q)}\
\\[10pt]
& \times&
\frac{S_{\rm NS}(\frac{Q}{2}+\tau +\alpha_1-\alpha_2)S_{\rm NS}(\frac{Q}{2}+\tau +\alpha_3-\bar\alpha_4)}
{S_{\rm NS}(\frac{Q}{2}+\lambda +\alpha_1-\alpha_2)S_{\rm NS}(\frac{Q}{2}+\lambda +\alpha_3-\bar\alpha_4)}
\
\left\langle\tau\left|^{\rm\scriptscriptstyle N}_{\rm\scriptscriptstyle N}\right|\xi_s\right\rangle
\left\langle \xi_s\left|^{\rm\scriptscriptstyle N}_{\rm\scriptscriptstyle N}\right|\lambda\right\rangle,
\end{eqnarray*}
and
\begin{eqnarray*}
\theta_{\rm\scriptscriptstyle NR}(x|\alpha_s)
\overline{\theta_{\rm\scriptscriptstyle NR}(y|\alpha_s)}
& = &
\int\hskip -15pt\int\limits_{-i\infty\;\;\;\;\; }^{i\infty\;}\!\!\! \frac{d\tau}{i}\frac{d\lambda}{i}\
{\rm e}^{\pi x(Q+\tau-\alpha_1-\alpha_2) + \pi y(-\lambda+\alpha_1+\alpha_2-Q)}\
\\
& \times\ &
\frac{S_{\rm NS}(\frac{Q}{2}+\tau +\alpha_1-\alpha_2)S_{\rm NS}(\frac{Q}{2}+\tau +\alpha_3-\bar\alpha_4)}
{S_{\rm NS}(\frac{Q}{2}+\lambda +\alpha_1-\alpha_2)S_{\rm NS}(\frac{Q}{2}+\lambda +\alpha_3-\bar\alpha_4)}
\
\left\langle\tau\left|^{\rm\scriptscriptstyle R}_{\rm\scriptscriptstyle R}\right|\xi_s\right\rangle
\left\langle \xi_s\left|^{\rm\scriptscriptstyle R}_{\rm\scriptscriptstyle R}\right|\lambda\right\rangle,
\end{eqnarray*}
we get
\begin{eqnarray*}
&&
\hskip -2cm
\int\limits_{\alpha_s \in \frac{Q}{2}+i{\mathbb R}_+}
\hskip -17pt \frac{d\alpha_s}{8i}\
\left|S_{\rm NS}(2\alpha_s)\right|^2\,
\left[
\theta_{\rm\scriptscriptstyle NN}(x|\alpha_s)
{\overline \theta_{\rm\scriptscriptstyle NN}(y|\alpha_s)}
+
\theta_{\rm\scriptscriptstyle NR}(x|\alpha_s)
{\overline \theta_{\rm\scriptscriptstyle NR}(y|\alpha_s)}
\right]
\\
& = &
\int\hskip -15pt\int\limits_{-i\infty\;\;\;\;\; }^{i\infty\;}\!\!\! \frac{d\tau}{i}\frac{d\lambda}{i}\
{\rm e}^{\pi x(Q+\tau-\alpha_1-\alpha_2) + \pi y(-\lambda+\alpha_1+\alpha_2-Q)}\
\\
& \times &
\frac{S_{\rm NS}(\frac{Q}{2}+\tau +\alpha_1-\alpha_2)S_{\rm NS}(\frac{Q}{2}+\tau +\alpha_3-\bar\alpha_4)}
{S_{\rm NS}(\frac{Q}{2}+\lambda +\alpha_1-\alpha_2)S_{\rm NS}(\frac{Q}{2}+\lambda +\alpha_3-\bar\alpha_4)}
\\
& \times &
\int\limits_{-i\infty}^{i\infty}\! \frac{d\xi_s}{16i}\
\nu(\xi_s)
\Big(
\left\langle\tau\left|^{\rm\scriptscriptstyle N}_{\rm\scriptscriptstyle N}\right|\xi_s\right\rangle
\left\langle \xi_s\left|^{\rm\scriptscriptstyle N}_{\rm\scriptscriptstyle N}\right|\lambda\right\rangle
+
\left\langle\tau\left|^{\rm\scriptscriptstyle R}_{\rm\scriptscriptstyle R}\right|\xi_s\right\rangle
\left\langle \xi_s\left|^{\rm\scriptscriptstyle R}_{\rm\scriptscriptstyle R}\right|\lambda\right\rangle
\Big)
\; = \;
\delta(x-y),
\end{eqnarray*}
where we used the symmetry $\xi_s \to -\xi_s$ of the function
\[
\nu(\xi_s)
\Big(
\left\langle\tau\left|^{\rm\scriptscriptstyle N}_{\rm\scriptscriptstyle N}\right|\xi_s\right\rangle
\left\langle \xi_s\left|^{\rm\scriptscriptstyle N}_{\rm\scriptscriptstyle N}\right|\lambda\right\rangle
+
\left\langle\tau\left|^{\rm\scriptscriptstyle R}_{\rm\scriptscriptstyle R}\right|\xi_s\right\rangle
\left\langle \xi_s\left|^{\rm\scriptscriptstyle R}_{\rm\scriptscriptstyle R}\right|\lambda\right\rangle
\Big)
\]
to extend the $\xi_s$  integration over the entire imaginary axis and  applied
Eq.\ (\ref{completeness:1}). Repeating essentially the same calculation (and using Eq.\
(\ref{completeness:2}) for the off-diagonal elements) we eventually get
\begin{equation}
\label{Theta:s:completeness}
\int\limits_{\alpha_s \in \frac{Q}{2}+i{\mathbb R}_+}
\hskip -17pt \frac{d\alpha_s}{i}\
\Theta_s(x|\alpha_s)\Theta^\dag_s(y|\alpha_s)
=
\sigma_0\,\delta(x-y).
\end{equation}
Analogous orthogonality and completeness relations are satisfied by $\Theta_t(x|\alpha_t),$
\begin{equation}
\label{Theta:t:orthogonality}
\int\limits_{\mathbb R}\!dx\
\Theta^\dag_t(x|\alpha_t) \Theta_t(x|\alpha_t')
=
\sigma_0\ i\delta(\alpha_t-\alpha_t'),
\end{equation}
and
\begin{equation}
\label{Theta:t:completeness}
\int\limits_{\alpha_t \in \frac{Q}{2}+i{\mathbb R}_+}
\hskip -17pt \frac{d\alpha_t}{i}\
\Theta_t(x|\alpha_t)\Theta^\dag_t(y|\alpha_t)
=
\sigma_0\,\delta(x-y).
\end{equation}
Consequently:
\begin{eqnarray*}
&&
\hskip -.5cm
\int\limits_{\frac{Q}{2}+i{\mathbb R}_+}\hskip -12pt \frac{d\alpha_s}{i}\
\left|S_{\rm NS}(2\alpha_s)\right|^2\,
{\mathbf G}_{\alpha_s\alpha_t}\!\left[^{\alpha_3\:\alpha_2}_{\alpha_4\:\alpha_1}\right]
{\mathbf G}^\dag_{\alpha_s\alpha_t'}\!\left[^{\alpha_3\:\alpha_2}_{\alpha_4\:\alpha_1}\right]
\\
& = &
S_{\rm NS}(2\alpha_t)\overline{S_{\rm NS}(2\alpha_t')}
\\
& \times &
\left({\mathbf M}^{t}_{\alpha_t}\!\left[^{\alpha_3\:\alpha_2}_{\alpha_4\:\alpha_1}\right]\right)^\dag
\int\hskip -8pt\int\limits_{\hskip -5pt\mathbb R}\!\!dx dy\
\Theta^\dag_t(x|\alpha_t)
\left(
\int\limits_{\frac{Q}{2}+i{\mathbb R}_+}\hskip -12pt \frac{d\alpha_s}{i}\
\Theta_s(x|\alpha_s)
\Theta^\dag_s(y|\alpha_s)
\right)
\Theta_t(y|\alpha_t')\
{\mathbf M}^{t}_{\alpha_t'}\!\left[^{\alpha_3\:\alpha_2}_{\alpha_4\:\alpha_1}\right]
\\
& = &
S_{\rm NS}(2\alpha_t)\overline{S_{\rm NS}(2\alpha_t')}\
\left({\mathbf M}^{t}_{\alpha_t}\!\left[^{\alpha_3\:\alpha_2}_{\alpha_4\:\alpha_1}\right]\right)^\dag
\int\limits_{\mathbb R}\!dx\
\Theta^\dag_t(x|\alpha_t)
\Theta_t(x|\alpha_t')\
{\mathbf M}^{t}_{\alpha_t'}\!\left[^{\alpha_3\:\alpha_2}_{\alpha_4\:\alpha_1}\right]
\\
& = &
\left|S_{\rm NS}(2\alpha_t)\right|^2
\sigma_0\
i\delta(\alpha_t-\alpha'_t),
\end{eqnarray*}
where we used (\ref{Theta:s:completeness}), (\ref{Theta:t:orthogonality}) and unitarity of
\(
{\mathbf M}^{t}_{\alpha_t}\!\left[^{\alpha_3\:\alpha_2}_{\alpha_4\:\alpha_1}\right].
\)
In consequence the equality (\ref{conjecture}) implies validity of the bootstrap equation for the
four-point correlator of the NS super-primary fields.

\subsection{Symmetry properties}
Using (\ref{small:thetas}) one can work out an explicit expression for the matrix $\mathbf G.$ It is of the form:
\begin{eqnarray}
\nonumber
{\mathbf G}_{\alpha_s\alpha_t}\!\left[^{\alpha_3\:\alpha_2}_{\alpha_4\:\alpha_1}\right]^\imath{}_{\!\jmath}
& = &
\frac
{S_{\imath}(\alpha_t + \alpha_1-\bar\alpha_4)S_{\imath}(\alpha_t - \alpha_3+\bar\alpha_2)}
{S_{\imath}(\alpha_t + \alpha_1-\alpha_4)S_{\imath}(\alpha_t - \alpha_3+\alpha_2)}
\frac
{S_{\!\jmath}(\alpha_s + \alpha_3-\alpha_4)S_{\!\jmath}(\alpha_s - \alpha_1+\alpha_2)}
{S_{\!\jmath}(\alpha_s + \alpha_3-\bar\alpha_4)S_{\!\jmath}(\alpha_s - \alpha_1+\bar\alpha_2)}
\\
\label{G:explicit}
& \times &
\frac{\left|S_{\rm NS}(2\alpha_t)\right|^2}{4} \, \int\limits_{-i\infty}^{i\infty}\!\frac{d\tau}{i}\
{\mathbf J}_{\alpha_s\alpha_t}\!\left[^{\alpha_3\:\alpha_2}_{\alpha_4\:\alpha_1}\right]^\imath{}_{\!\jmath},
\end{eqnarray}
where the superscript $\imath = 1$ (subscript $\jmath = 1$) on the l.h.s.\ corresponds to the function $S_{\rm NS}$ on the r.h.s,
the superscript $\imath = 2$ (subscript $\jmath = 2$) on the l.h.s.\ corresponds to the function $S_{\rm R}$ on the r.h.s.\ and:
\begin{eqnarray*}
{\mathbf J}_{\alpha_s\alpha_t}\!\left[^{\alpha_3\:\alpha_2}_{\alpha_4\:\alpha_1}\right]^1{}_1
& = &
\frac
{S_{\rm NS}(\tau +\bar\alpha_1)S_{\rm NS}(\tau+\alpha_4 +\alpha_2 -\alpha_3)S_{\rm NS}(\tau +\alpha_1)S_{\rm NS}(\tau+\alpha_4 +\alpha_2 -\bar\alpha_3)}
{S_{\rm NS}(\tau +\alpha_4+\bar\alpha_t)S_{\rm NS}(\tau +\alpha_4+\alpha_t)S_{\rm NS}(\tau +\alpha_2+\bar\alpha_s)S_{\rm NS}(\tau +\alpha_2+\alpha_s)}
\\[10pt]
& + &
\frac
{S_{\rm R}(\tau +\bar\alpha_1)S_{\rm R}(\tau+\alpha_4 +\alpha_2 -\alpha_3)S_{\rm R}(\tau +\alpha_1)S_{\rm R}(\tau+\alpha_4 +\alpha_2 -\bar\alpha_3)}
{S_{\rm R}(\tau +\alpha_4+\bar\alpha_t)S_{\rm R}(\tau +\alpha_4+\alpha_t)S_{\rm R}(\tau +\alpha_2+\bar\alpha_s)S_{\rm R}(\tau +\alpha_2+\alpha_s)},
\\[14pt]
{\mathbf J}_{\alpha_s\alpha_t}\!\left[^{\alpha_3\:\alpha_2}_{\alpha_4\:\alpha_1}\right]^1{}_2
& = &
\frac
{S_{\rm NS}(\tau +\bar\alpha_1)S_{\rm NS}(\tau+\alpha_4 +\alpha_2 -\alpha_3)S_{\rm NS}(\tau +\alpha_1)S_{\rm NS}(\tau+\alpha_4 +\alpha_2 -\bar\alpha_3)}
{S_{\rm NS}(\tau +\alpha_4+\bar\alpha_t)S_{\rm NS}(\tau +\alpha_4+\alpha_t)S_{\rm R}(\tau +\alpha_2+\bar\alpha_s)S_{\rm R}(\tau +\alpha_2+\alpha_s)}
\\[10pt]
& - &
\frac
{S_{\rm R}(\tau +\bar\alpha_1)S_{\rm R}(\tau+\alpha_4 +\alpha_2 -\alpha_3)S_{\rm R}(\tau +\alpha_1)S_{\rm R}(\tau+\alpha_4 +\alpha_2 -\bar\alpha_3)}
{S_{\rm R}(\tau +\alpha_4+\bar\alpha_t)S_{\rm R}(\tau +\alpha_4+\alpha_t)S_{\rm NS}(\tau +\alpha_2+\bar\alpha_s)S_{\rm NS}(\tau +\alpha_2+\alpha_s)},
\\[14pt]
{\mathbf J}_{\alpha_s\alpha_t}\!\left[^{\alpha_3\:\alpha_2}_{\alpha_4\:\alpha_1}\right]^2{}_1
& = &
\frac
{S_{\rm NS}(\tau +\bar\alpha_1)S_{\rm NS}(\tau+\alpha_4 +\alpha_2 -\alpha_3)S_{\rm NS}(\tau +\alpha_1)S_{\rm NS}(\tau+\alpha_4 +\alpha_2 -\bar\alpha_3)}
{S_{\rm R}(\tau +\alpha_4+\bar\alpha_t)S_{\rm R}(\tau +\alpha_4+\alpha_t)S_{\rm NS}(\tau +\alpha_2+\bar\alpha_s)S_{\rm NS}(\tau +\alpha_2+\alpha_s)}
\\[10pt]
& - &
\frac
{S_{\rm R}(\tau +\bar\alpha_1)S_{\rm R}(\tau+\alpha_4 +\alpha_2 -\alpha_3)S_{\rm R}(\tau +\alpha_1)S_{\rm R}(\tau+\alpha_4 +\alpha_2 -\bar\alpha_3)}
{S_{\rm NS}(\tau +\alpha_4+\bar\alpha_t)S_{\rm NS}(\tau +\alpha_4+\alpha_t)S_{\rm R}(\tau +\alpha_2+\bar\alpha_s)S_{\rm R}(\tau +\alpha_2+\alpha_s)},
\\[14pt]
{\mathbf J}_{\alpha_s\alpha_t}\!\left[^{\alpha_3\:\alpha_2}_{\alpha_4\:\alpha_1}\right]^2{}_2
& = &
\frac
{S_{\rm NS}(\tau +\bar\alpha_1)S_{\rm NS}(\tau+\alpha_4 +\alpha_2 -\alpha_3)S_{\rm NS}(\tau +\alpha_1)S_{\rm NS}(\tau+\alpha_4 +\alpha_2 -\bar\alpha_3)}
{S_{\rm R}(\tau +\alpha_4+\bar\alpha_t)S_{\rm R}(\tau +\alpha_4+\alpha_t)S_{\rm R}(\tau +\alpha_2+\bar\alpha_s)S_{\rm R}(\tau +\alpha_2+\alpha_s)}
\\[10pt]
& + &
\frac
{S_{\rm R}(\tau +\bar\alpha_1)S_{\rm R}(\tau+\alpha_4 +\alpha_2 -\alpha_3)S_{\rm R}(\tau +\alpha_1)S_{\rm R}(\tau+\alpha_4 +\alpha_2 -\bar\alpha_3)}
{S_{\rm NS}(\tau +\alpha_4+\bar\alpha_t)S_{\rm NS}(\tau +\alpha_4+\alpha_t)S_{\rm NS}(\tau +\alpha_2+\bar\alpha_s)S_{\rm NS}(\tau +\alpha_2+\alpha_s)}.
\end{eqnarray*}
Multiplying elements of (\ref{G:explicit}) with the corresponding
elements of normalization factors $\sf N$ (see Eq.\ (\ref{F:definition})) we obtain an explicit expression
for the matrix $\mathbf F$:
\begin{eqnarray}
\nonumber
{\mathbf F}_{\alpha_s\alpha_t}\!\left[^{\alpha_3\:\alpha_2}_{\alpha_4\:\alpha_1}\right]^\imath{}_{\!\jmath}
& = &
\frac{
\Gamma_\imath(\bar\alpha_t +\bar\alpha_3-\alpha_2)
\Gamma_\imath(\bar\alpha_t +\alpha_3-\alpha_2)
\Gamma_\imath(\alpha_t +\bar\alpha_3-\alpha_2)
\Gamma_\imath(\alpha_t +\alpha_3-\alpha_2)
}
{
\Gamma_{\!\jmath}(\bar\alpha_s +\bar\alpha_1-\alpha_2)
\Gamma_{\!\jmath}(\bar\alpha_s +\alpha_1-\alpha_2)
\Gamma_{\!\jmath}(\alpha_s +\bar\alpha_1-\alpha_2)
\Gamma_{\!\jmath}(\alpha_s +\alpha_1-\alpha_2)
}
\\[4pt]
\nonumber
& \times &
\frac{
\Gamma_\imath(\bar\alpha_t +\bar\alpha_1-\bar\alpha_4)
\Gamma_\imath(\bar\alpha_t +\alpha_1-\bar\alpha_4)
\Gamma_\imath(\alpha_t +\bar\alpha_1-\bar\alpha_4)
\Gamma_\imath(\alpha_t +\alpha_1-\bar\alpha_4)
}
{
\Gamma_{\!\jmath}(\bar\alpha_s +\bar\alpha_3-\bar\alpha_4)
\Gamma_{\!\jmath}(\bar\alpha_s +\alpha_3-\bar\alpha_4)
\Gamma_{\!\jmath}(\alpha_s +\bar\alpha_3-\bar\alpha_4)
\Gamma_{\!\jmath}(\alpha_s +\alpha_3-\bar\alpha_4)
}
\\[4pt]
\label{F:explicit}
& \times &
\frac{\Gamma_{\rm NS}(2\alpha_s)\Gamma_{\rm NS}(2\bar\alpha_s)}
{4\,\Gamma_{\rm NS}(\alpha_t-\bar\alpha_t)\Gamma_{\rm NS}(\bar\alpha_t-\alpha_t)}
\, \int\limits_{-i\infty}^{i\infty}\!\frac{d\tau}{i}\
{\mathbf J}_{\alpha_s\alpha_t}\!\left[^{\alpha_3\:\alpha_2}_{\alpha_4\:\alpha_1}\right]^\imath{}_{\!\jmath}.
\end{eqnarray}
Notice that $\mathbf F$ is explicitly invariant with respect to conjugations $\alpha_i\to Q-\alpha_i$ for
$i = s,t,1,3.$ Employing (\ref{F:reflection}) we further get:
\[
{\mathbf F}_{\alpha_s\alpha_t}\!\left[^{\alpha_3\:\alpha_2}_{\alpha_4\:\alpha_1}\right]
=
{\mathbf F}_{\alpha_s\alpha_t}\!\left[^{\alpha_2\:\alpha_3}_{\bar\alpha_1\:\bar\alpha_4}\right].
\]
The matrix on the r.h.s.\ of this equation is explicitly invariant with respect to conjugations
$\alpha_2 \to Q-\alpha_2$ and $\alpha_4\to Q- \alpha_4.$ This property therefore holds also for
the matrix on the l.h.s. The matrix $\mathbf F$ thus depends only on the conformal weights and
enjoys an invariance with respect to exchange of its columns,
\begin{equation}
\label{exchage:of:colums}
{\mathbf F}_{\alpha_s\alpha_t}\!\left[^{\alpha_3\:\alpha_2}_{\alpha_4\:\alpha_1}\right]
=
{\mathbf F}_{\alpha_s\alpha_t}\!\left[^{\alpha_2\:\alpha_3}_{\alpha_1\:\alpha_4}\right].
\end{equation}
Finally, since the deformed hypergeometric function (\ref{deformedF:definition}) is  invariant
with respect to exchange of its first two arguments,
\[
{\mathbb F}(\alpha,\beta;\gamma;z) = {\mathbb F}(\beta,\alpha;\gamma,z),
\]
all the factors in the definition of the matrix
\(
{\mathbf F}_{\alpha_s\alpha_t}\!\left[^{\alpha_3\:\alpha_2}_{\alpha_4\:\alpha_1}\right]
\)
are invariant with respect to the simultaneous exchange $\alpha_1\leftrightarrow\alpha_3$ and
$\alpha_2\leftrightarrow\bar\alpha_4.$ We thus have
\[
{\mathbf F}_{\alpha_s\alpha_t}\!\left[^{\alpha_3\:\alpha_2}_{\alpha_4\:\alpha_1}\right]
=
{\mathbf F}_{\alpha_s\alpha_t}\!\left[^{\alpha_1\:\bar\alpha_4}_{\alpha_2\:\alpha_3}\right]
=
{\mathbf F}_{\alpha_s\alpha_t}\!\left[^{\alpha_1\:\alpha_4}_{\alpha_2\:\alpha_3}\right],
\]
and using (\ref{exchage:of:colums}) we finally conclude that the
matrix
\(
{\mathbf F}_{\alpha_s\alpha_t}\!\left[^{\alpha_3\:\alpha_2}_{\alpha_4\:\alpha_1}\right]
\)
is invariant with respect to exchange of its rows as well,
\begin{equation}
\label{exchage:of:rows}
{\mathbf F}_{\alpha_s\alpha_t}\!\left[^{\alpha_3\:\alpha_2}_{\alpha_4\:\alpha_1}\right]
=
{\mathbf F}_{\alpha_s\alpha_t}\!\left[^{\alpha_4\:\alpha_1}_{\alpha_3\:\alpha_2}\right].
\end{equation}
\subsection{Special values of the arguments}
\label{secial:values}
Let us discuss the limit $\alpha_2\to -b.$ This value corresponds to the degenerate field
$V_{-b},$ whose operator product expansion with an arbitrary primary field $V_{\alpha_1}$ decomposes
onto three conformal families \cite{Belavin:2007gz}:
\begin{equation}
\label{OPE}
V_{-b}\,V_{\alpha_1}
\
\in
\
\left[V_{\alpha_1-b}\right]_{\rm\scriptstyle ee}
+
\left[V_{\alpha_1}\right]_{\rm\scriptstyle oo}
+
\left[V_{\alpha_1+b}\right]_{\rm\scriptstyle ee}.
\end{equation}
This property is reflected by the fact that the conformal block
\(
{\cal F}_{\!\alpha_s}^{\rm e}\!\left[^{\alpha_3\: \alpha_2}_{\alpha_4\: \alpha_1}\right]\!(z)
\)
possesses a well defined limit $\alpha_2 \to -b$ if and only if $\alpha_2 = \alpha_1 \pm b,$
while
\(
\lim\limits_{\alpha_2\to -b}
{\cal F}_{\!\alpha_s}^{\rm o}\!\left[^{\alpha_3\: \alpha_2}_{\alpha_4\: \alpha_1}\right]\!(z)
\)
exists if and only if $\alpha_s = \alpha_1.$ The four--point correlation function of
the field $V_{-b}$ and the super-primary fields $V_{\alpha_1},\, V_{\alpha_3},\, V_{\alpha_4}$
with arbitrary $\alpha_1,\alpha_3,\alpha_4 \in \frac{Q}{2} + i {\mathbb R}_+$ can be thus
expressed through these three blocks. Since this property cannot depend on our choice of the decomposition
``channel'', the integral in the equation (\ref{Fusion:definition}) must descent for $\alpha_s = \alpha_1 \pm b, 0$
and $\alpha_2 \to -b$ to the sum containing just three terms.

Let us check this for the block
\(
{\cal F}_{\!\alpha_1\!+b}^{\rm e}\!\left[^{\alpha_3\: -b}_{\alpha_4\:\ \alpha_1}\right]\!(z).
\)
To this end we need to calculate the limits
\begin{equation}
\label{alpha2:to:minus:b}
\lim_{\alpha_2\to 0}
{\mathbf F}_{\alpha_1\!+b,\alpha_t}\!\left[^{\alpha_3\:\alpha_2}_{\alpha_4\:\alpha_1}\right]^{\imath}{}_1
=
\lim_{\alpha_2\to 0}
{\mathbf F}_{\alpha_1\!+b,\alpha_t}\!\left[^{\alpha_4\:\alpha_1}_{\alpha_3\:\alpha_2}\right]^{\imath}{}_1,
\hskip 1cm
\imath = 1,2.
\end{equation}

For $\alpha_s = \alpha_1-b$ the pre-factor
of the integral in
\(
{\mathbf F}_{\alpha_1\!+b,\alpha_t}\!\left[^{\alpha_4\:\alpha_1}_{\alpha_3\:\alpha_2}\right]^\imath{}_1
\)
contains a term $\Gamma_{\rm NS}^{-1}(\alpha_2+b),$
and thus (\ref{alpha2:to:minus:b}) vanishes unless there is a compensating, singular factor
coming form the integral in (\ref{F:explicit}).
As explained in (\cite{Ponsot:2000mt}, Lemma 3), such a  singular term can arise in the process of analytic
continuation of the integral
\begin{eqnarray*}
&&
\int\limits_{-i\infty}^{i\infty}\!\frac{d\tau}{i}\
{\mathbf J}_{\alpha_1\!+b,\alpha_t}\!\left[^{\alpha_4\:\alpha_1}_{\alpha_3\:\alpha_2}\right]^1{}_{\!1}
= I_{\rm NS}(\alpha_2) +  I_{\rm R}(\alpha_2),
\\
&&
I_{\rm NS}(\alpha_2)
=
\hskip -5pt
\int\limits_{-i\infty}^{i\infty}\!\frac{d\tau}{i}\
\frac
{S_{\rm NS}(\tau +\bar\alpha_2)S_{\rm NS}(\tau+\alpha_3 +\alpha_1 -\alpha_4)S_{\rm NS}(\tau +\alpha_2)S_{\rm NS}(\tau+\alpha_3 +\alpha_1 -\bar\alpha_4)}
{S_{\rm NS}(\tau +\alpha_3+\bar\alpha_t)S_{\rm NS}(\tau +\alpha_3+\alpha_t)S_{\rm NS}(\tau +Q-b)S_{\rm NS}(\tau +2\alpha_1+b)},
\\
&&
I_{\rm R}(\alpha_2)
=
\hskip -5pt
\int\limits_{-i\infty}^{i\infty}\!\frac{d\tau}{i}\
\frac
{S_{\rm R}(\tau +\bar\alpha_2)S_{\rm R}(\tau+\alpha_3 +\alpha_1 -\alpha_4)S_{\rm R}(\tau +\alpha_2)S_{\rm R}(\tau+\alpha_3 +\alpha_1 -\bar\alpha_4)}
{S_{\rm R}(\tau +\alpha_3+\bar\alpha_t)S_{\rm R}(\tau +\alpha_3+\alpha_t)S_{\rm R}(\tau +Q-b)S_{\rm R}(\tau +2\alpha_1+b)},
\end{eqnarray*}
to $\alpha_2 = -b$ if the integration contour gets ``pinched'' between two poles of the integrand; the
singular contribution is obtained by calculating the residue at any one of these poles. Such a pinching
occurs in $I_{\rm NS}(\alpha_2),$ where the pole at $\tau = b,$ coming from zero of the function
$S_{\rm NS}(\tau +Q-b)$ in the denominator and located to the right of the contour, ``collides'' in the limit $\alpha_2\to -b$
with a pole at $\tau = - \alpha_2$ from the term $S_{\rm NS}(\tau +\alpha_2)$ in the numerator, located to the left of the contour.
Calculating the residue we get:
\[
I^{(0)}_{\rm NS}(\alpha_2)
=
2\frac
{
S_{\rm NS}(Q-2\alpha_2)S_{\rm NS}(\alpha_1+\alpha_3-\alpha_4-\alpha_2)S_{\rm NS}(\alpha_1+\alpha_3-\bar\alpha_4-\alpha_2)
}
{
S_{\rm NS}(Q-b-\alpha_2)S_{\rm NS}(2\alpha_1+b-\alpha_2)S_{\rm NS}(\alpha_3+\alpha_t-\alpha_2)S_{\rm NS}(\alpha_3+\bar\alpha_t-\alpha_2)
},
\]
and consequently:
\begin{eqnarray*}
&&
\hskip -.5cm
\lim_{\alpha_2\to -b}
{\mathbf F}_{\alpha_1+b,\alpha_t}\!\left[^{\alpha_3\:\alpha_2}_{\alpha_4\:\alpha_1}\right]^1{}_1
=
\frac{\Gamma_{\rm NS}(2\bar\alpha_1-2b)}{\Gamma_{\rm NS}(2\bar\alpha_1)}
\frac{\Gamma_{\rm NS}(\bar\alpha_4+\bar\alpha_t-\alpha_1)
\Gamma_{\rm NS}(\bar\alpha_4+\alpha_t-\alpha_1)}
{
\Gamma_{\rm NS}(\bar\alpha_4+\bar\alpha_3-\alpha_1-b)
\Gamma_{\rm NS}(\bar\alpha_4+\alpha_3-\alpha_1-b)}
\\[4pt]
& \times &
\frac
{
\Gamma_{\rm NS}(\alpha_4+\bar\alpha_t-\alpha_1)
\Gamma_{\rm NS}(\alpha_4+\alpha_t-\alpha_1)
}
{
\Gamma_{\rm NS}(\alpha_4+\bar\alpha_3-\alpha_1-b)
\Gamma_{\rm NS}(\alpha_4+\alpha_3-\alpha_1-b)
}
\\[4pt]
& \times &
\lim_{\alpha_2\to 0}\
\frac{
\Gamma_{\rm NS}(\alpha_3+\alpha_2-\alpha_t)\Gamma_{\rm NS}(\alpha_t-\alpha_3+\alpha_2)
\Gamma_{\rm NS}(\alpha_3+\alpha_2-\bar\alpha_t)\Gamma_{\rm NS}(\bar\alpha_t-\alpha_3+\alpha_2)
}
{
2\,\Gamma_{\rm NS}(Q) \Gamma_{\rm NS}(2\alpha_2)\Gamma_{\rm NS}(2\alpha_t -Q)\Gamma_{\rm NS}(Q-2\alpha_t)
}.
\end{eqnarray*}
Functions in the last line possess poles which move as we change $\alpha_2,$
pinching the contour of integration over $\alpha_t$ (which is originally the semi-line $\frac{Q}{2} + i{\mathbb R}_+$).
We get two pairs of colliding poles:
the poles of $\Gamma_{\rm NS}(\alpha_t-\alpha_3+\alpha_2)$ at $\alpha_t = \alpha_3 -\alpha_2$ and $\alpha_3 - \alpha_2 - 2b$
(to the left of the
contour) collide with the poles of $\Gamma_{\rm NS}(\alpha_3+\alpha_2-\alpha_t)$
at $\alpha_t = \alpha_3 + \alpha_2 + 2b$ and $\alpha_t = \alpha_3 + \alpha_2$  (to the right of the contour).
Calculating residues of the colliding poles we get:
\[
\lim_{\alpha_2\to -b}
\hskip -10pt
\int\limits_{\frac{Q}{2} + i{\mathbb R}_+}
\hskip -10pt \frac{d\alpha_t}{i}\
{\cal F}_{\alpha_t}^{e}\!\left[^{\alpha_1\: \alpha_2}_{\alpha_4\: \alpha_3}\right]\!(1-x)
\,
{\mathbf F}_{\alpha_1\!+b,\alpha_t}\!\left[^{\alpha_3\: \alpha_2}_{\alpha_4\: \alpha_1}\right]^{1}{}_1
=
\sum\limits_{s=\pm 1}
{\mathbf F}_{+,s}\left[^{\alpha_3\: -b}_{\alpha_4\:\ \alpha_1}\right]
{\cal F}_{\alpha_3+s\,b}^{e}\!\left[^{\alpha_1\: -b}_{\alpha_4\:\ \alpha_3}\right]\!(1-x),
\]
where
\begin{eqnarray}
\label{F:plus:plus}
{\mathbf F}_{+,+}\left[^{\alpha_3\: -b}_{\alpha_4\:\ \alpha_1}\right]
& = &
\frac{\Gamma_{\rm NS}(2\bar\alpha_1-2b)}{\Gamma_{\rm NS}(2\bar\alpha_1)}
\frac{\Gamma_{\rm NS}(Q-2\alpha_3)}{\Gamma_{\rm NS}(Q-2\alpha_3+2b)}
\\
\nonumber
&\times &
\frac{\Gamma_{\rm NS}(\alpha_4 +\alpha_3 -\alpha_1 + b)\Gamma_{\rm NS}(\bar\alpha_4 +\alpha_3 -\alpha_1 + b)}
{\Gamma_{\rm NS}(\alpha_4 +\alpha_3 -\alpha_1 - b)\Gamma_{\rm NS}(\bar\alpha_4 +\alpha_3 -\alpha_1 - b)},
\end{eqnarray}
and
\begin{eqnarray}
\label{F:plus:minus}
{\mathbf F}_{+,-}\left[^{\alpha_3\: -b}_{\alpha_4\:\ \alpha_1}\right]
& = &
\frac{\Gamma_{\rm NS}(2\bar\alpha_1-2b)}{\Gamma_{\rm NS}(2\bar\alpha_1)}
\frac{\Gamma_{\rm NS}(2\alpha_3-Q)}{\Gamma_{\rm NS}(2\alpha_3-Q+2b)}
\\
\nonumber
&\times &
\frac{\Gamma_{\rm NS}(\alpha_4 +\bar\alpha_3 -\alpha_1 + b)\Gamma_{\rm NS}(\bar\alpha_4 +\bar\alpha_3 -\alpha_1 + b)}
{\Gamma_{\rm NS}(\alpha_4 +\bar\alpha_3 -\alpha_1 - b)\Gamma_{\rm NS}(\bar\alpha_4 +\bar\alpha_3 -\alpha_1 - b)}.
\end{eqnarray}
The case
\[
\lim_{\alpha_2\to 0}
{\mathbf F}_{\alpha_1\!+b,\alpha_t}\!\left[^{\alpha_3\:\alpha_2}_{\alpha_4\:\alpha_1}\right]^{2}{}_1
=
\lim_{\alpha_2\to 0}
{\mathbf F}_{\alpha_1\!+b,\alpha_t}\!\left[^{\alpha_4\:\alpha_1}_{\alpha_3\:\alpha_2}\right]^{2}{}_1
\]
is similar. Calculating the residue of the colliding pole we get:
\begin{eqnarray*}
&&
\hskip -5mm
\int\limits_{-i\infty}^{i\infty}\!\frac{d\tau}{i}\
{\mathbf J}_{\alpha_1\!+b,\alpha_t}\!\left[^{\alpha_4\:\alpha_1}_{\alpha_3\:\alpha_2}\right]^2{}_{\!1}
\\
&&
=
2\frac
{
S_{\rm NS}(Q-2\alpha_2)S_{\rm NS}(\alpha_1+\alpha_3-\alpha_4-\alpha_2)S_{\rm NS}(\alpha_1+\alpha_3-\bar\alpha_4-\alpha_2)
}
{
S_{\rm NS}(Q-b-\alpha_2)S_{\rm NS}(2\alpha_1+b-\alpha_2)S_{\rm R}(\alpha_3+\alpha_t-\alpha_2)S_{\rm R}(\alpha_3+\bar\alpha_t-\alpha_2)
}
+ {\rm\ regular},
\end{eqnarray*}
and
\begin{eqnarray*}
&&
\hskip -.5cm
\lim_{\alpha_2\to -b}
{\mathbf F}_{\alpha_1+b,\alpha_t}\!\left[^{\alpha_3\:\alpha_2}_{\alpha_4\:\alpha_1}\right]^2{}_1
=
\frac{\Gamma_{\rm NS}(2\bar\alpha_1-2b)}{\Gamma_{\rm NS}(2\bar\alpha_1)}
\frac{
\Gamma_{\rm R}(\bar\alpha_4+\bar\alpha_t-\alpha_1)
\Gamma_{\rm R}(\bar\alpha_4+\alpha_t-\alpha_1)}{
\Gamma_{\rm NS}(\bar\alpha_4+\bar\alpha_3-\alpha_1-b)
\Gamma_{\rm NS}(\bar\alpha_4+\alpha_3-\alpha_1-b)}
\\[4pt]
& \times &
\frac
{
\Gamma_{\rm R}(\alpha_4+\bar\alpha_t-\alpha_1)
\Gamma_{\rm R}(\alpha_4+\alpha_t-\alpha_1)
}
{
\Gamma_{\rm NS}(\alpha_4+\bar\alpha_3-\alpha_1-b)
\Gamma_{\rm NS}(\alpha_4+\alpha_3-\alpha_1-b)
}
\\[4pt]
& \times &
\lim_{\alpha_2\to 0}\
\frac{
\Gamma_{\rm R}(\alpha_3+\alpha_2-\alpha_t)\Gamma_{\rm R}(\alpha_t-\alpha_3+\alpha_2)
\Gamma_{\rm R}(\alpha_3+\alpha_2-\bar\alpha_t)\Gamma_{\rm R}(\bar\alpha_t-\alpha_3+\alpha_2)
}
{
2\,\Gamma_{\rm NS}(Q) \Gamma_{\rm NS}(2\alpha_2)\Gamma_{\rm NS}(2\alpha_t -Q)\Gamma_{\rm NS}(Q-2\alpha_t)
}.
\end{eqnarray*}
This time in the last line there is only one pair of poles pinching the $\alpha_t$ contour: a pole
of the function $\Gamma_{\rm R}(\alpha_t-\alpha_3+\alpha_2)$ located at $\alpha_t = \alpha_3-\alpha_2-b$
collides in the limit $\alpha_2\to -b$ with a pole of the function $\Gamma_{\rm R}(\alpha_3+\alpha_2-\alpha_t)$ at
$\alpha_t = \alpha_3+\alpha_2 + b.$ Calculating the residue we get:
\[
\lim_{\alpha_2\to -b}
\hskip -10pt
\int\limits_{\frac{Q}{2} + i{\mathbb R}_+}
\hskip -10pt \frac{d\alpha_t}{i}\
{\cal F}_{\alpha_t}^{o}\!\left[^{\alpha_1\: \alpha_2}_{\alpha_4\: \alpha_3}\right]\!(1-x)
\,
{\mathbf F}_{\alpha_1\!+b,\alpha_3}\!\left[^{\alpha_3\: \alpha_2}_{\alpha_4\: \alpha_1}\right]^{2}{}_1
=
{\mathbf F}_{+,0}\left[^{\alpha_3\: -b}_{\alpha_4\:\ \alpha_1}\right]
{\cal F}_{\alpha_3}^{o}\!\left[^{\alpha_1\: -b}_{\alpha_4\:\ \alpha_3}\right]\!(1-x),
\]
where
\begin{eqnarray}
\nonumber
{\mathbf F}_{+,0}\left[^{\alpha_3\: -b}_{\alpha_4\:\ \alpha_1}\right]
& = &
c_0\,\frac{\Gamma_{\rm NS}(2\bar\alpha_1-2b)}{\Gamma_{\rm NS}(2\bar\alpha_1)}
\frac{\Gamma_{\rm R}(2\alpha_3 -Q-b)\Gamma_{\rm R}(Q-2\alpha_3-b)}{\Gamma_{\rm NS}(2\alpha_3-Q)\Gamma_{\rm NS}(Q-2\alpha_3)}\,
\times
\\
\label{F:plus:zero}
\nonumber
& \times &
\frac
{
\Gamma_{\rm R}(\bar\alpha_4+\bar\alpha_3-\alpha_1)
\Gamma_{\rm R}(\bar\alpha_4+\alpha_3-\alpha_1)
}
{
\Gamma_{\rm NS}(\bar\alpha_4+\bar\alpha_3-\alpha_1-b)
\Gamma_{\rm NS}(\bar\alpha_4+\alpha_3-\alpha_1-b)
}
\\
\nonumber
&\times &
\frac{
\Gamma_{\rm R}(\alpha_4+\bar\alpha_3-\alpha_1)
\Gamma_{\rm R}(\alpha_4+\alpha_3-\alpha_1)
}
{
\Gamma_{\rm NS}(\alpha_4+\bar\alpha_3-\alpha_1-b)
\Gamma_{\rm NS}(\alpha_4+\alpha_3-\alpha_1-b)
},
\end{eqnarray}
and where
\[
c_0 = \lim_{\alpha_2\to -b}
\frac{\Gamma_{\rm R}(\alpha_2)\Gamma_{\rm R}(2\alpha_2+b)}
{\Gamma_{\rm NS}(\alpha_2+b)\Gamma_{\rm NS}(2\alpha_2)}
=
2\cos\left(\frac{bQ}{2}\right)\frac{\Gamma(bQ)}{\Gamma\left(\frac{bQ}{2}\right)}\,b^{-\frac{bQ}{2}}.
\]
One can check in the same way that
the fusion equations for ${\cal F}_{\alpha_1}^{o}\!\left[^{\alpha_3\: -b}_{\alpha_4\:\ \alpha_1}\right]\!(z)$
as well as  for ${\cal F}_{\alpha_1\!-b}^{e}\!\left[^{\alpha_3\: -b}_{\alpha_4\:\ \alpha_1}\right]\!(z)$
also contain only three terms, proportional to
${\cal F}_{\alpha_3\!\pm b}^{e}\!\left[^{\alpha_1\: -b}_{\alpha_4\:\ \alpha_3}\right]\!(1-z)$
and
${\cal F}_{\alpha_3}^{0}\!\left[^{\alpha_1\: -b}_{\alpha_4\:\ \alpha_3}\right]\!(1-z)$
(the algebra involved becomes more elaborate --- in the case of
${\cal F}_{\alpha_1}^{o}$
one gets from the $\mathbf J$ integral two contributions from the colliding poles and in the
case of ${\cal F}_{\alpha_1\!-b}^{e}$
we have three pair of poles pinching the $\tau$ contour --- and we shall not present these calculations here).

As a final check let us calculate for
(\ref{F:plus:minus}), (\ref{F:plus:plus}) and (\ref{F:plus:zero})
the corresponding elements of the $\mathbf G$ matrix (\ref{G:definition}).
Using (\ref{F:definition}),  (\ref{normalization:N}) and (\ref{G:shift}) we get:
\begin{eqnarray}
\nonumber
{\mathbf G}_{+,-}\!\left[^{\alpha_3\: -b}_{\alpha_4\:\ \alpha_1}\right]
& = &
\frac{S_{\rm NS}(2\alpha_3-2b)}{S_{\rm NS}(2\alpha_3)}
\frac{S_{\rm NS}(\alpha_1+\alpha_3-\alpha_4+b)}{S_{\rm NS}(\alpha_1+\alpha_3-\alpha_4-b)}
\\[4pt]
\label{G:plus:minus}
& = &
\frac{\cos\frac{\pi b}{2}(\alpha_1+\alpha_3-\alpha_4)\,\sin\frac{\pi b}{2}(\alpha_1+\alpha_3-\alpha_4-b)}
{\cos\pi b\!\left(\alpha_3-\frac{b}{2}\right)\,\sin\pi b(\alpha_3-b)},
\\[6pt]
\nonumber
{\mathbf G}_{+,+}\!\left[^{\alpha_3\: -b}_{\alpha_4\:\ \alpha_1}\right]
& = &
\frac{S_{\rm NS}(2\alpha_3-Q)}{S_{\rm NS}(2\alpha_3-Q+2b)}
\frac{S_{\rm NS}(\alpha_1+\alpha_4-\alpha_3+b)}{S_{\rm NS}(\alpha_1+\alpha_4-\alpha_3-b)}
\\[4pt]
\label{G:plus:plus}
& = &
-\frac{\cos\frac{\pi b}{2}(\alpha_1+\alpha_4-\alpha_3)\,\sin\frac{\pi b}{2}(\alpha_1+\alpha_4-\alpha_3-b)}
{\cos\pi b\!\left(\alpha_3-\frac{b}{2}\right)\,\sin\pi b\alpha_3},
\end{eqnarray}
and
\begin{eqnarray}
\nonumber
{\mathbf G}_{+,0}\!\left[^{\alpha_3\: -b}_{\alpha_4\:\ \alpha_1}\right]
& = &
\frac{S_{\rm NS}(2\alpha_3)S_{\rm NS}(2\bar\alpha_3)}{S_{\rm R}(2\alpha_3+b)S_{\rm R}(2\bar\alpha_3+b)}
\frac{S_{\rm NS}(\alpha_1+\alpha_3-\alpha_4+b)S_{\rm NS}(\alpha_1+\alpha_4-\alpha_3+b)}
{S_{\rm R}(\alpha_1+\alpha_3-\alpha_4)S_{\rm R}(\alpha_1+\alpha_4-\alpha_3)}
\\[4pt]
\label{G:plus:zero}
& = &
\frac{\cos\frac{\pi b}{2}(\alpha_1+\alpha_3-\alpha_4)\cos\frac{\pi b}{2}(\alpha_1+\alpha_4-\alpha_3)}
{\sin\pi b\alpha_3\,\sin\pi b(\alpha_3-b)}.
\end{eqnarray}
It is reassuring to compare (\ref{G:plus:minus} -- \ref{G:plus:zero}) with the
results of \cite{Belavin:2007gz}, where
${\cal F}_{\alpha_1\!\pm b}^{e}\!\left[^{\alpha_3\: -b}_{\alpha_4\:\ \alpha_1}\right]\!(z)$
and
${\cal F}_{\alpha_1}^{o}\!\left[^{\alpha_3\: -b}_{\alpha_4\:\ \alpha_1}\right]\!(z)$
have been calculated (in the double integral representation of Dotsenko and Fatteev \cite{Dotsenko:1984nm})
by solving the corresponding null vector decoupling equation
(see especially Appendix C of \cite{Belavin:2007gz}).

\section{Special functions related to the Barnes gamma}
\label{special:functions}
\subsection{Definitions and main properties}
For $\Re\,x > 0$ the Barnes double gamma function has an integral
representation of the form:
\[
\log\,\Gamma_b(x)
\; = \;
\int\limits_{0}^{\infty}\frac{dt}{t}
\left[
\frac{{\rm e}^{- x t} - {\rm e}^{- {Q \over 2}t}}
{\left(1-{\rm e}^{- tb}\right)\left(1-{\rm e}^{- t/b}\right)}
-
\frac{\left({\textstyle{Q\over 2}}-x\right)^2}{2{\rm e}^{t}}
-
\frac{{\textstyle{Q\over 2}}-x}{t}
\right].
\]
It satisfies functional relations
\begin{eqnarray}
\nonumber
\Gamma_b(x+b)
& = &
\frac{\sqrt{2\pi}\,b^{bx-\frac12}}{\Gamma(bx)}\Gamma_b(x),
\\[-6pt]
\label{Gamma:b:shift}
\\[-6pt]
\nonumber
\Gamma_b\left(x + b^{-1}\right)
& = &
\frac{\sqrt{2\pi}\,b^{-\frac{x}{b}+\frac12}}{\Gamma(\frac{x}{b})}\Gamma_b(x),
\end{eqnarray}
and can be analytically continued to the whole complex $x$ plane as a
meromorphic function with no zeroes and with poles located at $ x = -m b - n{1\over b},\; m, n \in
{\mathbb N}.$ Relations (\ref{Gamma:b:shift}) allow to calculate residues of these poles
in terms of $\Gamma_b(Q);$ for instance for $x \to 0:$
\[
\Gamma_b(x) \; = \; \frac{\Gamma_b(Q)}{2\pi x} + {\cal O}(1).
\]
It is  convenient to introduce
\begin{eqnarray}
\label{otherspecial:b}
\Upsilon_b(x) & = & \frac{1}{\Gamma_b(x)\Gamma_b(Q-x)},
\hskip 5mm
S_b(x) = \frac{\Gamma_b(x)}{\Gamma_b(Q-x)},
\hskip 5mm
G_b(x) = {\rm e}^{-\frac{i\pi}{2}x(Q-x)} S_b(x),
\end{eqnarray}
and, borrowing the notation from \cite{Fukuda:2002bv}, to denote:
\begin{eqnarray}
\nonumber
\Gamma_{\rm NS}(x)
& = &
\Gamma_b\left(\frac{x}{2}\right)\Gamma_b\left(\frac{x+Q}{2}\right),
\hskip 1.1cm
\Gamma_{\rm R}(x)
=
\Gamma_b\left(\frac{x+b}{2}\right)\Gamma_b\left(\frac{x+b^{-1}}{2}\right),
\\[-5pt]
\label{susyspecial:defs}
\\[-5pt]
\nonumber
\Upsilon_{\rm NS}(x)
& = &
\Upsilon_b\left(\frac{x}{2}\right)\Upsilon_b\left(\frac{x+Q}{2}\right),
\hskip 1cm
\Upsilon_{\rm R}(x)
=
\Upsilon_b\left(\frac{x+b}{2}\right)\Upsilon_b\left(\frac{x+b^{-1}}{2}\right),
\end{eqnarray}
etc.

Using relations (\ref{Gamma:b:shift}) and definitions
(\ref{otherspecial:b}),\ (\ref{susyspecial:defs}) one can easily establish
basic properties of these functions. They include:
\subsubsection*{Relations between $S$ and $G$ functions}
\vskip -20pt
\begin{eqnarray*}
G_{\rm NS}(x) & = &
\zeta_0\,
{\rm e}^{- \frac{i\pi}{4}x(Q-x)}S_{\rm NS}(x),
\\[10pt]
G_{\rm R}(x) & = &
{\rm e}^{- \frac{i\pi}{4}}\zeta_0\,
{\rm e}^{- \frac{i\pi}{4}x(Q-x)}S_{\rm R}(x),
\end{eqnarray*}
where
\[
\zeta_0 \;\ = \;\ {\rm e}^{- \frac{i\pi Q^2}{8}}.
\]
\subsubsection*{Shift relations}
\vskip -20pt
\begin{eqnarray}
\nonumber
S_{\rm NS}(x+b^{\pm 1}) & = & 2\cos\left(\frac{\pi b^{\pm 1} x}{2}\right)\,S_{\rm R}(x),
\hskip 1cm
S_{\rm R}(x+b^{\pm 1}) = 2\sin\left(\frac{\pi b^{\pm 1} x}{2}\right)\,S_{\rm NS}(x),
\\[-5pt]
\label{G:shift}
\\[-5pt]
\nonumber
G_{\rm NS}(x+b^{\pm 1}) & = & \left(1+{\rm e}^{i\pi b^{\pm 1} x}\right)G_{\rm R}(x),
\hskip .6cm
G_{\rm R}(x+b^{\pm 1}) = \left(1- {\rm e}^{i\pi b^{\pm 1} x}\right)G_{\rm NS}(x).
\end{eqnarray}
\subsubsection*{Reflection properties}
\vskip -20pt
\begin{eqnarray*}
S_{\rm NS}(x)S_{\rm NS}(Q-x) & = &
S_{\rm R}(x)S_{\rm R}(Q-x) = 1
\end{eqnarray*}
and consequently:
\begin{eqnarray}
\nonumber
G_{\rm NS}(x)G_{\rm NS}(Q-x) & = &
\zeta_0^2\,
{\rm e}^{- \frac{i\pi}{2}x(Q-x)},
\\[-5pt]
\label{G:reflections}
\\[-5pt]
\nonumber
G_{\rm R}(x)G_{\rm R}(Q-x) & = &
{\rm e}^{- \frac{i\pi}{2}}\zeta_0^2\,
{\rm e}^{- \frac{i\pi}{2}x(Q-x)}.
\end{eqnarray}
\subsubsection*{Asymptotic behavior}
\vskip -20pt
\begin{eqnarray*}
G_{\rm NS}(x)
& \to &
\left\{
\begin{array}{lll}
1,
&&
x \to +i\infty,
\\
\zeta_0^2\,{\rm e}^{- \frac{i\pi}{2}x(Q-x)},
&&
x \to -i\infty,
\end{array}
\right.
\\[10pt]
G_{\rm R}(x)
& \to &
\left\{
\begin{array}{lll}
1,
&&
x \to +i\infty,
\\
{\rm e}^{-\frac{i\pi}{2}}\zeta_0^2\,{\rm e}^{- \frac{i\pi}{2}x(Q-x)},
&&
x \to -i\infty.
\end{array}
\right.
\end{eqnarray*}
\subsubsection*{Zeroes and poles}
\vskip -20pt
\begin{eqnarray*}
S_{\rm NS}(x) & = & 0
\hskip 5mm
\Leftrightarrow
\hskip 5mm
x = Q+ mb+nb^{-1},
\hskip 5mm
m,n \in {\mathbb Z}_{\geq 0}, \;
m+n \in 2{\mathbb Z},
\\
S_{\rm R}(x) & = & 0
\hskip 5mm
\Leftrightarrow
\hskip 5mm
x = Q+ mb+nb^{-1},
\hskip 5mm
m,n \in {\mathbb Z}_{\geq 0}, \;
m+n \in 2{\mathbb Z} +1,
\\
S_{\rm NS}(x)^{-1} & = & 0
\hskip 5mm
\Leftrightarrow
\hskip 5mm
x = -mb-nb^{-1},
\hskip 5mm
m,n \in {\mathbb Z}_{\geq 0}, \;
m+n \in 2{\mathbb Z},
\\
S_{\rm R}(x)^{-1} & = & 0
\hskip 5mm
\Leftrightarrow
\hskip 5mm
x = -mb-nb^{-1},
\hskip 5mm
m,n \in {\mathbb Z}_{\geq 0}, \;
m+n \in 2{\mathbb Z} +1.
\end{eqnarray*}
\subsubsection*{Basic residues}
\vskip -20pt
\begin{eqnarray}
\label{residues}
\lim_{x\to 0}\ x\,S_{\rm NS}(x) & = & \frac{1}{\pi},
\hskip 1cm
\lim_{x\to 0}\ x\,G_{\rm NS}(x) =  \frac{1}{\pi}\zeta_0.
\end{eqnarray}

\subsection{Integral identities}
We shall derive  several identities satisfied by a hypergeometric-type integrals,
generalizing results of
\cite{Kashaev}.

\noindent
Let us denote:
\begin{eqnarray}
\nonumber
B_{\rm NS}^{(\pm)}(\alpha,\beta)
& = &
\hskip -3pt
\int\limits_{-i\infty}^{i\infty}\!\frac{d\tau}{i}\
{\rm e}^{i\pi \beta}
\left[
\frac{G_{\rm NS}(\tau + \alpha)}{G_{\rm NS}(\tau + Q -0^{+})}
\pm
\frac{G_{\rm R}(\tau + \alpha)}{G_{\rm R}(\tau + Q)}
\right],
\\[-5pt]
\label{Beta:def}
\\[-5pt]
\nonumber
B_{\rm R}^{(\pm)}(\alpha,\beta)
& = &
\hskip -3pt
\int\limits_{-i\infty}^{i\infty}\!\frac{d\tau}{i}\
{\rm e}^{i\pi \beta}
\left[
\frac{G_{\rm R}(\tau + \alpha)}{G_{\rm NS}(\tau + Q -0^{+})}
\pm
\frac{G_{\rm NS}(\tau + \alpha)}{G_{\rm R}(\tau + Q)}
\right].
\end{eqnarray}
Using (\ref{G:shift}) one derives an identities:
\begin{eqnarray}
\nonumber
\left(1+{\rm e}^{i\pi b(\alpha + \beta)}\right)
B_{\rm NS}^{(+)}(\alpha + b,\beta)
& = &
\left(1+{\rm e}^{i\pi b\alpha}\right)B_{\rm R}^{(+)}(\alpha,\beta),
\\[10pt]
\nonumber
\left(1-{\rm e}^{i\pi b(\alpha + \beta)}\right)B_{\rm R}^{(+)}(\alpha + b,\beta)
& = &
\left(1-{\rm e}^{i\pi b\alpha}\right)B_{\rm NS}^{(+)}(\alpha,\beta),
\\[10pt]
\label{Beta:shifts:1}
\left(1-{\rm e}^{i\pi b(\alpha + \beta)}\right)B_{\rm NS}^{(-)}(\alpha + b,\beta)
& = &
\left(1+{\rm e}^{i\pi b\alpha}\right)B_{\rm R}^{(-)}(\alpha,\beta),
\\[10pt]
\nonumber
\left(1+{\rm e}^{i\pi b(\alpha + \beta)}\right)B_{\rm R}^{(-)}(\alpha + b,\beta)
& = &
\left(1-{\rm e}^{i\pi b\alpha}\right)B_{\rm NS}^{(-)}(\alpha,\beta),
\end{eqnarray}
and
\begin{eqnarray}
\nonumber
\left(1+{\rm e}^{i\pi b(\alpha + \beta)}\right)B_{\rm NS}^{(+)}(\alpha,\beta+b)
& = &
\left(1+{\rm e}^{i\pi b\beta}\right)B_{\rm NS}^{(-)}(\alpha,\beta),
\\[10pt]
\nonumber
\left(1-{\rm e}^{i\pi b(\alpha + \beta)}\right)B_{\rm NS}^{(-)}(\alpha,\beta+b)
& = &
\left(1-{\rm e}^{i\pi b\beta}\right)B_{\rm NS}^{(+)}(\alpha,\beta),
\\[10pt]
\label{Beta:shifts:2}
\left(1-{\rm e}^{i\pi b(\alpha + \beta)}\right)B_{\rm R}^{(+)}(\alpha,\beta+b)
& = &
\left(1+{\rm e}^{i\pi b\beta}\right)B_{\rm R}^{(-)}(\alpha,\beta),
\\[10pt]
\nonumber
\left(1+{\rm e}^{i\pi b(\alpha + \beta)}\right)B_{\rm R}^{(-)}(\alpha,\beta + b)
& = &
\left(1-{\rm e}^{i\pi b\beta}\right)B_{\rm R}^{(+)}(\alpha,\beta).
\end{eqnarray}
For $b\in {\mathbb R} \setminus {\mathbb Q}$ equations (\ref{Beta:shifts:1}), (\ref{Beta:shifts:2}),
together with their counterparts obtained by substituting $b \to b^{-1},$ determine the functions
involved up to $\alpha$ and $\beta$ independent factors,
\begin{eqnarray}
\nonumber
B_{\rm NS}^{(+)}(\alpha,\beta)
& = &
C_{\rm NS}^{(+)}\,\frac{G_{\rm NS}(\alpha)G_{\rm NS}(\beta)}{G_{\rm NS}(\alpha+\beta)},
\hskip 1cm
B_{\rm NS}^{(-)}(\alpha,\beta)
\;\ = \;\
C_{\rm NS}^{(-)}\,\frac{G_{\rm NS}(\alpha)G_{\rm R}(\beta)}{G_{\rm R}(\alpha+\beta)},
\\[-2pt]
\label{Beta:ntl}
\\[-2pt]
\nonumber
B_{\rm R}^{(+)}(\alpha,\beta)
& = &
C_{\rm R}^{(+)}\,\frac{G_{\rm R}(\alpha)G_{\rm NS}(\beta)}{G_{\rm R}(\alpha+\beta)},
\hskip 1.2 cm
B_{\rm R}^{(-)}(\alpha,\beta)
\; = \;
C_{\rm R}^{(-)}\,\frac{G_{\rm R}(\alpha)G_{\rm R}(\beta)}{G_{\rm NS}(\alpha+\beta)}.
\end{eqnarray}
The factors $C_{\rm NS,R}^{(\pm)}$ can be computed by explicitly calculating all the
functions appearing in (\ref{Beta:ntl}) at special values of $\alpha;$ since:
\begin{eqnarray*}
B_{\rm R}^{(\pm)}\left(b^{-1},\beta\right)
& = &
\int\limits_{-i\infty}^{i\infty}\!\frac{d\tau}{i}\
{\rm e}^{i\pi\tau\beta}
\left[
\frac{1}{1-{\rm e}^{i\pi b\tau}}
\pm
\frac{1}{1+{\rm e}^{i\pi b\tau}}
\right]
\; = \;
\frac{2}{ib(1\mp {\rm e}^{i\pi b^{-1}\beta})},
\end{eqnarray*}
and
\begin{eqnarray*}
B_{\rm NS}^{(+)}\left(b^{-1}-b,\beta\right)
& = &
\int\limits_{-i\infty}^{i\infty}\!\frac{d\tau}{i}
\frac{{\rm e}^{i\pi\tau\beta}}
{\left(1-{\rm e}^{i\pi b\tau}\right)\left(1+q^{-1}{\rm e}^{i\pi b\tau}\right)}
\; = \;
\frac{2}{ib(1+q)}\frac{q + {\rm e}^{i\pi Q\beta}}{1-{\rm e}^{2\pi i b^{-1}\beta}},
\\
B_{\rm NS}^{(-)}\left(b^{-1}-b,\beta\right)
& = &
\int\limits_{-i\infty}^{i\infty}\!\frac{d\tau}{i}
\frac{{\rm e}^{i\pi\tau\beta}}
{\left(1+{\rm e}^{i\pi b\tau}\right)\left(1-q^{-1}{\rm e}^{i\pi b\tau}\right)}
\; = \;
\frac{2}{ib(1+q)}\frac{q{\rm e}^{i\pi b^{-1}\beta} + {\rm e}^{i\pi b\beta}}{1-{\rm e}^{2\pi i b^{-1}\beta}},
\end{eqnarray*}
we  get a ``supersymmetric'', integral analog of the Ramanujan summation formula:
\begin{eqnarray}
\nonumber
\int\limits_{-i\infty}^{i\infty}\!\frac{d\tau}{i}\
{\rm e}^{i\pi\tau\beta}
\frac{G_{\rm NS}(\tau + \alpha)}{G_{\rm NS}(\tau + Q-0^+)}
& = &
{\rm e}^{\frac{i\pi Q^2}{8}}
G_{\rm NS}(\alpha)
\left[
\frac{G_{\rm NS}(\beta)}{G_{\rm NS}(\alpha+\beta)}
+
\frac{G_{\rm R}(\beta)}{G_{\rm R}(\alpha+\beta)}
\right],
\\
\nonumber
\int\limits_{-i\infty}^{i\infty}\!\frac{d\tau}{i}\
{\rm e}^{i\pi\tau\beta}
\frac{G_{\rm R}(\tau + \alpha)}{G_{\rm R}(\tau + Q)}
& = &
{\rm e}^{\frac{i\pi Q^2}{8}}
G_{\rm NS}(\alpha)
\left[
\frac{G_{\rm NS}(\beta)}{G_{\rm NS}(\alpha+\beta)}
-
\frac{G_{\rm R}(\beta)}{G_{\rm R}(\alpha+\beta)}
\right],
\\[-6pt]
\label{SIRS}
\\[-6pt]
\nonumber
\int\limits_{-i\infty}^{i\infty}\!\frac{d\tau}{i}\
{\rm e}^{i\pi\tau\beta}
\frac{G_{\rm R}(\tau + \alpha)}{G_{\rm NS}(\tau + Q-0^+)}
& = &
{\rm e}^{\frac{i\pi Q^2}{8}}
G_{\rm R}(\alpha)
\left[
\frac{G_{\rm NS}(\beta)}{G_{\rm R}(\alpha+\beta)}
+
\frac{G_{\rm R}(\beta)}{G_{\rm NS}(\alpha+\beta)}
\right],
\\
\nonumber
\int\limits_{-i\infty}^{i\infty}\!\frac{d\tau}{i}\
{\rm e}^{i\pi\tau\beta}
\frac{G_{\rm NS}(\tau + \alpha)}{G_{\rm R}(\tau + Q)}
& = &
{\rm e}^{\frac{i\pi Q^2}{8}}
G_{\rm R}(\alpha)
\left[
\frac{G_{\rm NS}(\beta)}{G_{\rm R}(\alpha+\beta)}
-
\frac{G_{\rm R}(\beta)}{G_{\rm NS}(\alpha+\beta)}
\right].
\end{eqnarray}
Formulae (\ref{SIRS}) allow to derive an analog of the ${}_1{\hskip -1pt}F_2$ Heine transformation.
Let us introduce a shorthand notation:
\begin{eqnarray*}
\left[
\hskip -3pt
\begin{array}{c}
{\rm NN}
\\[-4pt]
{\rm NN}
\end{array}
\hskip -3pt
\right]\!
(\alpha,\beta;\gamma;z)
& = &
\int\limits_{-i\infty}^{i\infty}\!\frac{d\tau}{i}\
{\rm e}^{i\pi\tau z}
\frac{G_{\rm NS}(\tau + \alpha) G_{\rm NS}(\tau + \beta)}{G_{\rm NS}(\tau + \gamma-0^+) G_{\rm NS}(\tau + Q-0^+)},
\\
\left[
\hskip -3pt
\begin{array}{c}
{\rm RR}
\\[-4pt]
{\rm NN}
\end{array}
\hskip -3pt
\right]\!
(\alpha,\beta;\gamma;z)
& = &
\int\limits_{-i\infty}^{i\infty}\!\frac{d\tau}{i}\
{\rm e}^{i\pi\tau z}
\frac{G_{\rm R}(\tau + \alpha) G_{\rm R}(\tau + \beta)}{G_{\rm NS}(\tau + \gamma-0^+) G_{\rm NS}(\tau + Q-0^+)},
\\
 & \vdots &
\\
\left(
\hskip -3pt
\begin{array}{c}
\alpha  \\[-4pt] \beta
\end{array}
\hskip -3pt
\right)_{\rm NS}
& = &
\frac{G_{\rm NS}(\alpha)}{G_{\rm NS}(\beta)},
\hskip 1cm
\left(
\hskip -3pt
\begin{array}{c}
\alpha  \\[-4pt] \beta
\end{array}
\hskip -3pt
\right)_{\rm R}
=
\frac{G_{\rm R}(\alpha)}{G_{\rm R}(\beta)}.
\end{eqnarray*}
Using (\ref{SIRS}) to express a ratio of the first function from the numerator and the first function
from the denominator as an integral, changing the order of integration and using (\ref{SIRS}) again we get:
\begin{eqnarray}
\nonumber
\left(
\left[
\hskip -3pt
\begin{array}{c}
{\rm NN}
\\[-4pt]
{\rm NN}
\end{array}
\hskip -3pt
\right]
+
\left[
\hskip -3pt
\begin{array}{c}
{\rm RR}
\\[-4pt]
{\rm RR}
\end{array}
\hskip -3pt
\right]
\right)
(\alpha,\beta;\gamma;z)
& = &
\left(
\hskip -3pt
\begin{array}{c}
 \beta \\[-4pt] \gamma -\alpha
\end{array}
\hskip -3pt
\right)_{\rm NS}
\left(
\left[
\hskip -3pt
\begin{array}{c}
{\rm NN}
\\[-4pt]
{\rm NN}
\end{array}
\hskip -3pt
\right]
+
\left[
\hskip -3pt
\begin{array}{c}
{\rm RR}
\\[-4pt]
{\rm RR}
\end{array}
\hskip -3pt
\right]
\right)
(z,\gamma-\alpha;z+\beta;\alpha),
\\[10pt]
\nonumber
\left(
\left[
\hskip -3pt
\begin{array}{c}
{\rm NN}
\\[-4pt]
{\rm NN}
\end{array}
\hskip -3pt
\right]
-
\left[
\hskip -3pt
\begin{array}{c}
{\rm RR}
\\[-4pt]
{\rm RR}
\end{array}
\hskip -3pt
\right]
\right)
(\alpha,\beta;\gamma;z)
& = &
\left(
\hskip -3pt
\begin{array}{c} \beta \\[-4pt] \gamma -\alpha \end{array}
\hskip -3pt
\right)_{\rm NS}
\left(
\left[
\hskip -3pt
\begin{array}{c} {\rm RN}  \\[-4pt] {\rm RN} \end{array}
\hskip -3pt
\right]
+
\left[
\hskip -3pt
\begin{array}{c} {\rm NR}  \\[-4pt] {\rm NR} \end{array}
\hskip -3pt
\right]
\right)
(z,\gamma-\alpha;z+\beta;\alpha),
\\[-6pt]
\label{Heine}
\\[-6pt]
\nonumber
\left(
\left[
\hskip -3pt
\begin{array}{c} {\rm RN}  \\[-4pt] {\rm RN} \end{array}
\hskip -3pt
\right]
-
\left[
\hskip -3pt
\begin{array}{c} {\rm NR}  \\[-4pt] {\rm NR} \end{array}
\hskip -3pt
\right]
\right)
(\alpha,\beta;\gamma;z)
& = &
\left(
\hskip -3pt
\begin{array}{c} \beta \\[-4pt] \gamma -\alpha \end{array}
\hskip -3pt
\right)_{\rm NS}
\left(
\left[
\hskip -3pt
\begin{array}{c} {\rm RN}  \\[-4pt] {\rm RN} \end{array}
\hskip -3pt
\right]
-
\left[
\hskip -3pt
\begin{array}{c} {\rm NR}  \\[-4pt] {\rm NR} \end{array}
\hskip -3pt
\right]
\right)
(z,\gamma-\alpha;z+\beta;\alpha),
\\[10pt]
\nonumber
\left(
\left[
\hskip -3pt
\begin{array}{c} {\rm RN}  \\[-4pt] {\rm RN} \end{array}
\hskip -3pt
\right]
+
\left[
\hskip -3pt
\begin{array}{c} {\rm NR}  \\[-4pt] {\rm NR} \end{array}
\hskip -3pt
\right]
\right)
(\alpha,\beta;\gamma;z)
& = &
\left(
\hskip -3pt
\begin{array}{c} \beta \\[-4pt] \gamma -\alpha \end{array}
\hskip -3pt
\right)_{\rm NS}
\left(
\left[
\hskip -3pt
\begin{array}{c} {\rm NN} \\[-4pt] {\rm NN} \end{array}
\hskip -3pt
\right]
-
\left[
\hskip -3pt
\begin{array}{c} {\rm RR} \\[-4pt] {\rm RR} \end{array}
\hskip -3pt
\right]
\right)
(z,\gamma-\alpha;z+\beta;\alpha),
\end{eqnarray}
plus twelve similar formulae with the other combinations of the $G_{\rm NS,R}$ functions.

Combining (three times) formulae (\ref{Heine}) with an exchange of the first two arguments of the involved functions one arrives
at the ``supersymmetric'' integral analogues of the Euler-Heine transformations. Four out of sixteen
formulae of this type read:
\begin{eqnarray}
\nonumber
&&
\hskip -1cm
\left(
\left[
\hskip -3pt
\begin{array}{c} {\rm NN} \\[-4pt] {\rm NN} \end{array}
\hskip -3pt
\right]
+
\left[
\hskip -3pt
\begin{array}{c} {\rm RR} \\[-4pt] {\rm RR} \end{array}
\hskip -3pt
\right]
\right)
(\alpha,\beta;\gamma;z)
\\[4pt]
\nonumber
& = &
\left(
\hskip -3pt
\begin{array}{c}
\alpha\\[-4pt] \gamma - \beta
\end{array}
\hskip -3pt
\right)_{\rm NS}
\left(
\hskip -3pt
\begin{array}{c}
 z \\[-4pt] z-A
\end{array}
\hskip -3pt
\right)_{\rm NS}
\left(
\hskip -3pt
\begin{array}{c}
 \beta \\[-4pt] \gamma -\alpha
\end{array}
\hskip -3pt
\right)_{\rm NS}
\left(
\left[
\hskip -3pt
\begin{array}{c} {\rm NN} \\[-4pt] {\rm NN} \end{array}
\hskip -3pt
\right]
+
\left[
\hskip -3pt
\begin{array}{c} {\rm RR} \\[-4pt] {\rm RR} \end{array}
\hskip -3pt
\right]
\right)
(\gamma-\beta,\gamma-\alpha;\gamma;z-A),
\\[10pt]
\nonumber
&&
\hskip -1cm
\left(
\left[
\hskip -3pt
\begin{array}{c} {\rm NN} \\[-4pt] {\rm NN} \end{array}
\hskip -3pt
\right]
-
\left[
\hskip -3pt
\begin{array}{c} {\rm RR} \\[-4pt] {\rm RR} \end{array}
\hskip -3pt
\right]
\right)
(\alpha,\beta;\gamma;z)
\\[4pt]
\nonumber
& = &
\left(
\hskip -3pt
\begin{array}{c}
 \alpha \\[-4pt] \gamma -\beta
\end{array}
\hskip -3pt
\right)_{\rm NS}
\left(
\hskip -3pt
\begin{array}{c}
 z \\[-4pt] z-A
\end{array}
\hskip -3pt
\right)_{\rm R}
\left(
\hskip -3pt
\begin{array}{c}
\beta\\[-4pt] \gamma - \alpha
\end{array}
\hskip -3pt
\right)_{\rm NS}
\left(
\left[
\hskip -3pt
\begin{array}{c} {\rm NN} \\[-4pt] {\rm NN} \end{array}
\hskip -3pt
\right]
-
\left[
\hskip -3pt
\begin{array}{c} {\rm RR} \\[-4pt] {\rm RR} \end{array}
\hskip -3pt
\right]
\right)
(\gamma-\beta,\gamma-\alpha;\gamma;z-A),
\\[10pt]
\label{Euler:Heine}
&&
\hskip -1cm
\left(
\left[
\hskip -3pt
\begin{array}{c}
{\rm RN} \\[-4pt] {\rm RN}
\end{array}
\hskip -3pt
\right]
-
\left[
\hskip -3pt
\begin{array}{c}
{\rm NR} \\[-4pt] {\rm NR}
\end{array}
\hskip -3pt
\right]
\right)
(\alpha,\beta;\gamma;z)
\\[4pt]
\nonumber
& = &
\left(
\hskip -3pt
\begin{array}{c}
\alpha \\[-4pt] \gamma - \beta
\end{array}
\hskip -3pt
\right)_{\rm R}
\left(
\hskip -3pt
\begin{array}{c}
 z \\[-4pt] z-A
\end{array}
\hskip -3pt
\right)_{\rm R}
\left(
\hskip -3pt
\begin{array}{c}
 \beta \\[-4pt] \gamma -\alpha
\end{array}
\hskip -3pt
\right)_{\rm NS}
\left(
\left[
\hskip -3pt
\begin{array}{c}
{\rm RN} \\[-4pt] {\rm RN}
\end{array}
\hskip -3pt
\right]
-
\left[
\hskip -3pt
\begin{array}{c}
{\rm NR} \\[-4pt] {\rm NR}
\end{array}
\hskip -3pt
\right]
\right)
(\gamma-\beta,\gamma-\alpha;\gamma;z-A),
\\[10pt]
\nonumber
&&
\hskip -1cm
\left(
\left[
\hskip -3pt
\begin{array}{c}
{\rm RN} \\[-4pt] {\rm RN}
\end{array}
\hskip -3pt
\right]
+
\left[
\hskip -3pt
\begin{array}{c}
{\rm NR} \\[-4pt] {\rm NR}
\end{array}
\hskip -3pt
\right]
\right)
(\alpha,\beta;\gamma;z)
\\[4pt]
\nonumber
& = &
\left(
\hskip -3pt
\begin{array}{c}
\alpha \\[-4pt] \gamma - \beta
\end{array}
\hskip -3pt
\right)_{\rm R}
\left(
\hskip -3pt
\begin{array}{c}
 z \\[-4pt] z-A
\end{array}
\hskip -3pt
\right)_{\rm NS}
\left(
\hskip -3pt
\begin{array}{c}
 \beta \\[-4pt] \gamma -\alpha
\end{array}
\hskip -3pt
\right)_{\rm NS}
\left(
\left[
\hskip -3pt
\begin{array}{c}
{\rm RN} \\[-4pt] {\rm RN}
\end{array}
\hskip -3pt
\right]
-
\left[
\hskip -3pt
\begin{array}{c}
{\rm NR} \\[-4pt] {\rm NR}
\end{array}
\hskip -3pt
\right]
\right)
(\gamma-\beta,\gamma-\alpha;\gamma;z-A),
\end{eqnarray}
where
\[
A = \gamma-\alpha -\beta.
\]
Reflection properties of the $G$ functions (\ref{G:reflections}) allow to write
\[
\frac{G_{\rm NS}(z)}{G_{\rm NS}(z-A)}
=
{\rm e}^{-\frac{i\pi}{2}A(Q+A-2z)}
\frac{G_{\rm NS}(Q+A-z)}{G_{\rm NS}(Q-z)},
\]
and similarly for the other combinations of R/NS. Using this type of relations,
formulae (\ref{SIRS}) and taking Fourier transform of equations (\ref{Euler:Heine})
one obtains a set of integral identities analogous  to the Saalsch\"utz summation formula. In particular
one gets:
\begin{eqnarray}
\label{SLong:1}
\nonumber
&&
\hskip -.5cm
\frac{1}{i}\!\!
\int\limits_{-i\infty}^{i\infty}\!\! d\tau\
{\rm e}^{i\pi\tau Q}
\left[
\frac{G_{\rm NS}(\tau + \alpha) G_{\rm NS}(\tau + \beta)G_{\rm NS}(\tau + \gamma)}
{G_{\rm NS}(\tau +\delta)G_{\rm NS}(\tau + \alpha+\beta+\gamma-\delta +Q) G_{\rm NS}(\tau + Q)}
\right.
\\[5pt]
  &&
\hskip 1.5cm
\left.
+\;
\frac{G_{\rm R}(\tau + \alpha) G_{\rm R}(\tau + \beta)G_{\rm R}(\tau + \gamma)}
{G_{\rm R}(\tau +\delta)G_{\rm R}(\tau + \alpha+\beta+\gamma-\delta +Q) G_{\rm R}(\tau + Q)}
\right]
\\[10pt]
  \nonumber
& = &
2\zeta_0^{-3}\,{\rm e}^{\frac{i\pi}{2}\delta(Q-\delta)}
\frac{G_{\rm NS}(\alpha)G_{\rm NS}(\beta)G_{\rm NS}(\gamma)G_{\rm NS}(Q+\alpha-\delta)G_{\rm NS}(Q+\beta-\delta)G_{\rm NS}(Q+\gamma-\delta)}
{G_{\rm NS}(Q+\alpha+\beta-\delta)G_{\rm NS}(Q+\alpha+\gamma-\delta)G_{\rm NS}(Q+\beta+\gamma-\delta)},
\end{eqnarray}
and
\begin{eqnarray}
 \label{SLong:2}
\nonumber
&&
\hskip -.5cm
\frac{1}{i}\!\!
\int\limits_{-i\infty}^{i\infty}\!\! d\tau\
{\rm e}^{i\pi\tau Q}
\left[
\frac{G_{\rm NS}(\tau + \alpha) G_{\rm NS}(\tau + \beta)G_{\rm R}(\tau + \gamma)}
{G_{\rm R}(\tau +\delta)G_{\rm NS}(\tau + \alpha+\beta+\gamma-\delta  +Q) G_{\rm NS}(\tau + Q)}
\right.
\\[5pt]
  &&
\hskip 1.5cm
\left.
+\;
\frac{G_{\rm R}(\tau + \alpha) G_{\rm R}(\tau + \beta)G_{\rm NS}(\tau + \gamma)}
{G_{\rm NS}(\tau +\delta)G_{\rm R}(\tau + \alpha+\beta+\gamma-\delta  +Q) G_{\rm R}(\tau + Q)}
\right]
\\[10pt]
  \nonumber
& = &
2i\zeta_0^{-3}\,{\rm e}^{\frac{i\pi}{2}\delta(Q-\delta)}
\frac{G_{\rm NS}(\alpha)G_{\rm NS}(\beta)G_{\rm R}(\gamma)G_{\rm R}(Q+\alpha-\delta)G_{\rm R}(Q+\beta-\delta)G_{\rm NS}(Q+\gamma-\delta)}
{G_{\rm R}(Q+\alpha+\beta-\delta)G_{\rm NS}(Q+\alpha+\gamma-\delta)G_{\rm NS}(Q+\beta+\gamma-\delta)}.
\end{eqnarray}
Taking in these equations the limit $\gamma \to i\infty$ we
finally obtain the formulae:
\begin{eqnarray}
\label{Sshort:1}
\nonumber
&&
\hskip -2cm
\frac{1}{i}\!\!
\int\limits_{-i\infty}^{i\infty}\!\! d\tau\
{\rm e}^{i\pi\tau Q}
\left[
\frac{G_{\rm NS}(\tau + \alpha) G_{\rm NS}(\tau + \beta)}
{G_{\rm NS}(\tau +\delta)G_{\rm NS}(\tau + Q)}
+
\frac{G_{\rm R}(\tau + \alpha) G_{\rm R}(\tau + \beta)}
{G_{\rm R}(\tau +\delta) G_{\rm R}(\tau + Q)}
\right]
\\[10pt]
& = &
2\zeta_0^{-3}\,{\rm e}^{\frac{i\pi}{2}\delta(Q-\delta)}
\frac{G_{\rm NS}(\alpha)G_{\rm NS}(\beta)G_{\rm NS}(Q+\alpha-\delta)G_{\rm NS}(Q+\beta-\delta)}
{G_{\rm NS}(Q+\alpha+\beta-\delta)}
\end{eqnarray}
and
\begin{eqnarray}
 \label{SShort:2}
\nonumber
&&
\hskip -2cm
\frac{1}{i}\!\!
\int\limits_{-i\infty}^{i\infty}\!\! d\tau\
{\rm e}^{i\pi\tau Q}
\left[
\frac{G_{\rm NS}(\tau + \alpha) G_{\rm NS}(\tau + \beta)}
{G_{\rm R}(\tau +\delta)G_{\rm NS}(\tau + Q)}
+
\frac{G_{\rm R}(\tau + \alpha) G_{\rm R}(\tau + \beta)}
{G_{\rm NS}(\tau +\delta)G_{\rm R}(\tau + Q)}
\right]
\\[10pt]
& = &
2i\zeta_0^{-3}\,{\rm e}^{\frac{i\pi}{2}\delta(Q-\delta)}
\frac{G_{\rm NS}(\alpha)G_{\rm NS}(\beta)G_{\rm R}(Q+\alpha-\delta)G_{\rm R}(Q+\beta-\delta)}
{G_{\rm R}(Q+\alpha+\beta-\delta)},
\end{eqnarray}
which will be the main tool in the proof of the orthogonality relation presented in the next subsection.

\subsection{Orthogonality and completeness relations}
Define for $\xi \in i{\mathbb R}_+$
\begin{eqnarray*}
\left\langle \tau\left|^{\rm\scriptscriptstyle N}_{\rm\scriptscriptstyle N}\right|\xi\right\rangle
& = &
\frac{1}{S_{\rm NS}(Q+\tau+\xi-0^+)S_{\rm NS}(Q+\tau-\xi-0^+)}
\;\ = \;\
\frac{S_{\rm NS}(\xi-\tau)}{S_{\rm NS}(Q+\tau+\xi-0^+)},
\\[6pt]
\left\langle \tau\left|^{\rm\scriptscriptstyle R}_{\rm\scriptscriptstyle N}\right|\xi\right\rangle
& = &
\frac{1}{S_{\rm NS}(Q+\tau+\xi-0^+)S_{\rm R}(Q+\tau-\xi)}
\;\ = \;\
\frac{S_{\rm R}(\xi-\tau)}{S_{\rm NS}(Q+\xi+\tau-0^+)},
\end{eqnarray*}
etc.\ and
\begin{eqnarray*}
\left\langle \tau\left|^{\rm\scriptscriptstyle N}_{\rm\scriptscriptstyle N}\right|\xi\right\rangle^{(\epsilon)}
& = &
\frac{1}{S_{\rm NS}(Q+\tau+\xi-\epsilon)S_{\rm NS}(Q+\tau-\xi-\epsilon)}
\;\ = \;\
\frac{S_{\rm NS}(\xi-\tau+\epsilon)}{S_{\rm NS}(Q+\tau+\xi-\epsilon)}.
\end{eqnarray*}

\subsubsection*{Orthogonality}
Using the relation between $S$ and $G$ functions as well as the  formulae
(\ref{Sshort:1}) we get:
\begin{eqnarray*}
&&
\hskip -1cm
\int\limits_{-i\infty}^{i\infty}\!\frac{d\tau}{i}\
\left[
\left\langle \xi_1\left|^{\rm\scriptscriptstyle N}_{\rm\scriptscriptstyle N}\right|\tau\right\rangle^{(\epsilon)}
\left\langle \tau\left|^{\rm\scriptscriptstyle N}_{\rm\scriptscriptstyle N}\right|\xi_2\right\rangle^{(\epsilon)}
+
\left\langle \xi_1\left|^{\rm\scriptscriptstyle R}_{\rm\scriptscriptstyle R}\right|\tau\right\rangle^{(\epsilon)}
\left\langle \tau\left|^{\rm\scriptscriptstyle R}_{\rm\scriptscriptstyle R}\right|\xi_2\right\rangle^{(\epsilon)}
\right]
\\
& = &
{\rm e}^{\frac{i\pi}{2}\left(\xi_2^2-\xi_1^2\right)}
\int\limits_{-i\infty}^{i\infty}\!\frac{d\tau}{i}\
{\rm e}^{i\pi Q\tau}
\left[
\frac{G_{\rm NS}(\tau+\xi_1+\epsilon)G_{\rm NS}(\tau-\xi_1+\epsilon)}
{G_{\rm NS}(Q+\tau+\xi_2-\epsilon)G_{\rm NS}(Q+\tau-\xi_2-\epsilon)}
\right.
\\
&&
\left.
\hskip 4cm
+\;
\frac{G_{\rm R}(\tau+\xi_1+\epsilon)G_{\rm R}(\tau-\xi_1+\epsilon)}
{G_{\rm R}(Q+\tau+\xi_2-\epsilon)G_{\rm R}(Q+\tau-\xi_2-\epsilon)}
\right]
\\[10pt]
& = &
2\zeta_0^{-3}
{\rm e}^{\frac{i\pi}{2}\left(\xi_2^2-\xi_1^2\right) -2i\pi\xi_2^2}
\frac{G_{\rm NS}(2\epsilon+ \xi_+)G_{\rm NS}(2\epsilon- \xi_+)
G_{\rm NS}(2\epsilon+\xi_-)G_{\rm NS}(2\epsilon-\xi_-)}
{G_{\rm NS}(4\epsilon)},
\end{eqnarray*}
where $\xi_{\pm} = \xi_2\pm\xi_1.$ In view of (\ref{residues})
the r.h.s.\ vanishes in the limit $\epsilon \to 0$ unless $\xi_- = 0$ (we cannot have $\xi_+ = 0$ since
$\Im\,\xi_i > 0,\ i = 1,2$). Consequently:
\begin{eqnarray}
\nonumber
&&
\hskip -2cm
\int\limits_{-i\infty}^{i\infty}\!\frac{d\tau}{i}\
\left[
\left\langle \xi_1\left|^{\rm\scriptscriptstyle N}_{\rm\scriptscriptstyle N}\right|\tau\right\rangle
\left\langle \tau\left|^{\rm\scriptscriptstyle N}_{\rm\scriptscriptstyle N}\right|\xi_2\right\rangle
+
\left\langle \xi_1\left|^{\rm\scriptscriptstyle R}_{\rm\scriptscriptstyle R}\right|\tau\right\rangle
\left\langle \tau\left|^{\rm\scriptscriptstyle R}_{\rm\scriptscriptstyle R}\right|\xi_2\right\rangle
\right]
\\
\label{orthogonality:1}
& = &
4\zeta_0^{-2}
{\rm e}^{\frac{i\pi}{2}\left(\xi_2^2-\xi_1^2\right) -2i\pi\xi_2^2}\
G_{\rm NS}(\xi_{+})G_{\rm NS}(-\xi_{+})\
\lim_{\epsilon\to 0}\frac{2\epsilon}{\pi\left(4\epsilon^2 - \xi_-^2\right)}
\\
\nonumber
& = &
4\frac{S_{\rm NS}(2\xi_2)}{S_{\rm NS}(2\xi_2+Q)}\ \delta(p_2-p_1)
=
\frac{4}{\nu(\xi_2)}\ \delta(p_2-p_1),
\end{eqnarray}
where $\xi_i = ip_i,\ i = 1,2$ and
\[
\nu(\xi) = -4\sin\pi b \xi\sin \pi b^{-1}\xi.
\]

Similarly:
\begin{eqnarray*}
&&
\hskip -1cm
\int\limits_{-i\infty}^{i\infty}\!\frac{d\tau}{i}\
\left[
\left\langle \xi_1\left|^{\rm\scriptscriptstyle N}_{\rm\scriptscriptstyle N}\right|\tau\right\rangle^{(\epsilon)}
\left\langle \tau\left|^{\rm\scriptscriptstyle R}_{\rm\scriptscriptstyle R}\right|\xi_2\right\rangle^{(\epsilon)}
-
\left\langle \xi_1\left|^{\rm\scriptscriptstyle R}_{\rm\scriptscriptstyle R}\right|\tau\right\rangle^{(\epsilon)}
\left\langle \tau\left|^{\rm\scriptscriptstyle N}_{\rm\scriptscriptstyle N}\right|\xi_2\right\rangle^{(\epsilon)}
\right]
\\
& = &
-i{\rm e}^{\frac{i\pi}{2}\left(\xi_2^2-\xi_1^2\right)}
\int\limits_{-i\infty}^{i\infty}\!\frac{d\tau}{i}\
{\rm e}^{i\pi Q\tau}
\left[
\frac{G_{\rm NS}(\tau+\xi_1+\epsilon)G_{\rm NS}(\tau-\xi_1+\epsilon)}
{G_{\rm R}(Q+\tau+\xi_2-\epsilon)G_{\rm R}(Q+\tau-\xi_2-\epsilon)}
\right.
\\
&&
\left.
\hskip 4.5cm
+\;
\frac{G_{\rm R}(\tau+\xi_1+\epsilon)G_{\rm R}(\tau-\xi_1+\epsilon)}
{G_{\rm NS}(Q+\tau+\xi_2-\epsilon)G_{\rm NS}(Q+\tau-\xi_2-\epsilon)}
\right]
\\[10pt]
& = &
-2i\zeta_0^{-3}
{\rm e}^{\frac{i\pi}{2}\left(\xi_2^2-\xi_1^2\right) -2i\pi\xi_2^2}
\frac{G_{\rm R}(2\epsilon+ \xi_+)G_{\rm R}(2\epsilon- \xi_+)
G_{\rm R}(2\epsilon+\xi_-)G_{\rm R}(2\epsilon-\xi_-)}
{G_{\rm NS}(4\epsilon)}.
\end{eqnarray*}
The function $G_{\rm R}(x)$ is regular in the vicinity of the imaginary axis
and taking the limit $\epsilon\to 0$ we have
for $\xi_1,\xi_2 \in i{\mathbb R}_+:$
\begin{equation}
\label{orthogonality:2}
\int\limits_{-i\infty}^{i\infty}\!\frac{d\tau}{i}\
\left[
\left\langle \xi_1\left|^{\rm\scriptscriptstyle N}_{\rm\scriptscriptstyle N}\right|\tau\right\rangle
\left\langle \tau\left|^{\rm\scriptscriptstyle R}_{\rm\scriptscriptstyle R}\right|\xi_2\right\rangle
-
\left\langle \xi_1\left|^{\rm\scriptscriptstyle R}_{\rm\scriptscriptstyle R}\right|\tau\right\rangle
\left\langle \tau\left|^{\rm\scriptscriptstyle N}_{\rm\scriptscriptstyle N}\right|\xi_2\right\rangle
\right]
=
0.
\end{equation}
\subsection*{Completeness}
Define
\[
\nu_\epsilon(\xi) \;\ = \;\ -4\sin\left(\pi b_\epsilon \xi\right)\sin\left(\pi b^{-1}_{\epsilon}\xi\right),
\hskip 5mm
b^{\pm1}_{\epsilon}\;\ = \;\ b^{\pm 1}-\epsilon,
\hskip 5mm
\lambda_{\epsilon} \;\ = \;\ \lambda+\epsilon,
\]
and consider  an integral:
\begin{eqnarray*}
&&
\hskip -.5cm
{\cal I}^\epsilon(\lambda,\rho)
=
\int\limits_{-i\infty}^{i\infty}\!\frac{d\tau}{i}\
\int\limits_{-i\infty}^{i\infty}\!\frac{d\xi}{i}\
\nu_{\epsilon}(\xi)
\left[
\left\langle \tau -\lambda_\epsilon\left|^{\rm\scriptscriptstyle N}_{\rm\scriptscriptstyle N}\right|\xi\right\rangle
\left\langle \xi\left|^{\rm\scriptscriptstyle N}_{\rm\scriptscriptstyle N}\right|\tau\right\rangle
+
\left\langle \tau -\lambda_\epsilon\left|^{\rm\scriptscriptstyle R}_{\rm\scriptscriptstyle R}\right|\xi\right\rangle
\left\langle \xi\left|^{\rm\scriptscriptstyle R}_{\rm\scriptscriptstyle R}\right|\tau\right\rangle
\right]
{\rm e}^{-i\pi\rho\tau}
\\[10pt]
& = &
\sum\limits_{k=1}^4
\frac{(-1)^k}{2}
\left[
\int\limits_{-i\infty}^{i\infty}\!\frac{du}{i}\
{\rm e}^{-\frac{i\pi u}{2}(\rho-\rho_k)}
\frac{S_{\rm NS}(u)}{S_{\rm NS}(u-\lambda_\epsilon+Q)}
\right]
\left[
\int\limits_{-i\infty}^{i\infty}\!\frac{dv}{i}\
{\rm e}^{\frac{i\pi v}{2}(\rho+\rho_k)}
\frac{S_{\rm NS}(v+\lambda_\epsilon)}{S_{\rm NS}(v+Q)}
\right]
\\[10pt]
& + &
\sum\limits_{k=1}^4
\frac{(-1)^k}{2}
\left[
\int\limits_{-i\infty}^{i\infty}\!\frac{du}{i}\
{\rm e}^{-\frac{i\pi u}{2}(\rho-\rho_k)}
\frac{S_{\rm R}(u)}{S_{\rm R}(u-\lambda_\epsilon+Q)}
\right]
\left[
\int\limits_{-i\infty}^{i\infty}\!\frac{dv}{i}\
{\rm e}^{\frac{i\pi v}{2}(\rho+\rho_k)}
\frac{S_{\rm R}(v+\lambda_\epsilon)}{S_{\rm R}(v+Q)}
\right],
\end{eqnarray*}
where
\(
u = \tau + \xi,\
v = \tau -\xi,
\)
and
\(
\rho_1 = -\rho_3 = b + b^{-1}-2\epsilon,\
\rho_2 = -\rho_4 = b-b^{-1}.
\)
It can be calculated by means of the formulae presented in the Appendix A and we get:
\[
{\cal I}^{\epsilon}(\lambda,\rho)
\;\ = \;\
{\cal I}_1^{\epsilon}(\lambda,\rho)
-
{\cal I}_2^{\epsilon}(\lambda,\rho)
+
{\cal I}_3^{\epsilon}(\lambda,\rho)
-
{\cal I}_4^{\epsilon}(\lambda,\rho)
\]
where
\begin{eqnarray*}
{\cal I}_1^{\epsilon}(\lambda,\rho)
& = &
2S^2_{\rm NS}(\lambda_\epsilon)
\frac{G_{\rm NS}\left(\frac{\rho-\lambda_\epsilon}{2}+\epsilon\right)}{G_{\rm NS}\left(\frac{\rho+\lambda_\epsilon}{2}+\epsilon\right)}
\frac{G_{\rm NS}\left(Q+\frac{\rho-\lambda_\epsilon}{2}-\epsilon\right)}{G_{\rm NS}\left(Q+\frac{\rho+\lambda_\epsilon}{2}-\epsilon\right)},
\\[10pt]
{\cal I}_2^{\epsilon}(\lambda,\rho)
& = &
2S^2_{\rm NS}(\lambda_\epsilon)
\frac{G_{\rm R}\left(b+\frac{\rho-\lambda_\epsilon}{2}\right)}{G_{\rm R}\left(b+\frac{\rho+\lambda_\epsilon}{2}\right)}
\frac{G_{\rm R}\left(b^{-1}+\frac{\rho-\lambda_\epsilon}{2}\right)}{G_{\rm R}\left(b^{-1}+\frac{\rho+\lambda_\epsilon}{2}\right)},
\\[10pt]
{\cal I}_3^{\epsilon}(\lambda,\rho)
& = &
2S^2_{\rm NS}(\lambda_\epsilon)
\frac{G_{\rm R}\left(\frac{\rho-\lambda_\epsilon}{2}+\epsilon\right)}{G_{\rm R}\left(\frac{\rho+\lambda_\epsilon}{2}+\epsilon\right)}
\frac{G_{\rm R}\left(Q+\frac{\rho-\lambda_\epsilon}{2}-\epsilon\right)}{G_{\rm R}\left(Q+\frac{\rho+\lambda_\epsilon}{2}-\epsilon\right)},
\\[10pt]
{\cal I}_4^{\epsilon}(\lambda,\rho)
& = &
2S^2_{\rm NS}(\lambda_\epsilon)
\frac{G_{\rm NS}\left(b+\frac{\rho-\lambda_\epsilon}{2}\right)}{G_{\rm NS}\left(b+\frac{\rho+\lambda_\epsilon}{2}\right)}
\frac{G_{\rm NS}\left(b^{-1}+\frac{\rho-\lambda_\epsilon}{2}\right)}{G_{\rm NS}\left(b^{-1}+\frac{\rho+\lambda_\epsilon}{2}\right)}.
\end{eqnarray*}

It is immediate to check with the help of  relations
(\ref{G:shift}) that outside of the singularities of the functions involved
$\lim\limits_{\epsilon\to 0}{\cal I}^{\epsilon}(\lambda,\rho) = 0.$ However, since
for $\epsilon\to 0$ some of the singularities approach the lines $\Im\,\rho = 0$ and $\Im\,\lambda = 0$
(the integration contours for the distribution ${\cal I}(\lambda,\rho)$), we have to be more careful.
The correct way of proceeding is analogous to the calculation in section \ref{secial:values}.

For $\varphi(\lambda,\rho)$ being a test function consider
\vskip -5pt
\begin{equation}
\label{distributional}
\varphi_i
\;\ = \;\
\lim_{\epsilon\to 0}
\int\limits_{-i\infty}^{i\infty}\!\frac{d\lambda}{i}\!
\int\limits_{-i\infty}^{i\infty}\!\frac{d\rho}{i}\
{\cal I}^{\epsilon}_i(\lambda,\rho)
\varphi(\lambda,\rho).
\end{equation}
${\cal I}_i(\lambda,\rho)$ have poles at imaginary $\lambda$
and $\rho$ axis.
From the form of ${\cal I}^{\epsilon}_2,\ {\cal I}^{\epsilon}_3$ and ${\cal I}^{\epsilon}_4$ it is clear that
all one has to do to define the limit $\epsilon \to 0$
of these terms is to deform the  contour of integration over $\lambda$ such that it avoids the singularity at $\lambda = -\epsilon.$
Since no poles pinching the integration contours appear, there are no contributions from the residues and
\[
\varphi_i
\;\ = \;\
\int\limits_{{\cal C}_\lambda}\!\frac{d\lambda}{i}\!
\int\limits_{{\cal C}_\rho}\!\frac{d\rho}{i}\
{\cal I}_i(\lambda,\rho)
\varphi(\lambda,\rho),
\hskip 1cm
i = 2,3,4,
\]
where ${\cal C}_\lambda$ and ${\cal C}_\rho$ denote the deformed contours.

The situation for ${\cal I}^{\epsilon}_1$ is different. In the complex $\rho$ plane a function
\(
G_{\rm NS}\left(\frac{\rho-\lambda_\epsilon}{2}+\epsilon\right)
\)
has a pole at $\rho = \rho_- = \lambda_\epsilon-2\epsilon = \lambda - \epsilon$ (to the left of the integration contour
$\rho \in i{\mathbb R})$, while a function
\(
G_{\rm NS}\left(Q+\frac{\rho+\lambda_\epsilon}{2}-\epsilon\right)^{-1}
\)
has a pole at $\rho = \rho_+ = -\lambda_\epsilon+ 2\epsilon = -\lambda + \epsilon$ (to the right of the integration contour).
For $\lambda, \epsilon  \to 0$ these poles collide. Choosing to deform the contour
past the pole at $\rho = \rho_-,$  taking into account (\ref{residues}) and
the relation
\begin{eqnarray*}
&&
\hskip -3cm
\zeta_0
\lim_{\epsilon\to 0}
\int\limits_{-i\infty}^{i\infty}\!\frac{d\lambda}{i}\
\frac{S_{\rm NS}^2(\lambda_\epsilon)G_{\rm NS}(Q-2\epsilon)}
{G_{\rm NS}(\lambda_{\epsilon})G_{\rm NS}(Q+\lambda_\epsilon -2\epsilon)} \varphi(\lambda,\lambda_\epsilon - 2\epsilon)
\\
& = &
\lim_{\epsilon\to 0}
\int\limits_{-i\infty}^{i\infty}\!\frac{d\lambda}{i}\
{\rm e}^{\frac{i\pi}{2}Q\lambda}\
\frac{S_{\rm NS}(\epsilon+\lambda)S_{\rm NS}(\epsilon-\lambda)}{S_{\rm NS}(2\epsilon)}\varphi(\lambda,\lambda-\epsilon)
\\
& = &
\lim_{\epsilon\to 0}
\int\limits_{-i\infty}^{i\infty}\!\frac{d\lambda}{i}\
{\rm e}^{\frac{i\pi}{2}Q\lambda}\
\frac{2\epsilon}{\pi\left(\epsilon^2-\lambda^2\right)}\varphi(\lambda,\lambda-\epsilon)
\;\ = \;\
2\varphi(0,0),
\end{eqnarray*}
we get
\begin{eqnarray*}
\varphi_1
& = &
\int\limits_{{\cal C}_\lambda}\!\frac{d\lambda}{i}\
\!\int\limits_{{\cal C}_\rho}\!\frac{d\rho}{i}\
{\cal I}_1(\lambda,\rho)
\varphi(\lambda,\rho)
+ 16\varphi(0,0),
\end{eqnarray*}
and finally
\begin{eqnarray*}
 \sum\limits_{k=1}^4\, \varphi_k
& = &
16\varphi(0,0)
+ \;
\int\limits_{{\cal C}_\lambda}\!\frac{d\lambda}{i}\
\int\limits_{{\cal C}_\rho}\frac{d\rho}{i}\
\sum\limits_{k=1}^k (-1)^{k-1}{\cal I}_k(\lambda,\rho)
\varphi(\lambda,\rho)
\\[10pt]
& = &
16\varphi(0,0)
+ \;
\int\limits_{{\cal C}_\lambda}\!\frac{d\lambda}{i}\
\int\limits_{{\cal C}_\rho}\frac{d\rho}{i}\
\left[\sum\limits_{k=1}^k (-1)^{k-1}{\cal I}_k(\lambda,\rho)\right]
\varphi(\lambda,\rho)
\;\ = \;\
16\varphi(0,0),
\end{eqnarray*}
what proves the equality
\[
\lim_{\epsilon\to 0}{\cal I}_1^{\epsilon}(\lambda,\rho)
\;\ = \;\ 16\delta(\lambda)\delta(\rho).
\]
Taking the inverse Fourier transform we have
\begin{equation}
 \label{completeness:1}
 \int\limits_{-i\infty}^{i\infty}\!\frac{d\xi}{i}\
\nu(\xi)
\Big(
\left\langle \eta -\lambda\left|^{\rm\scriptscriptstyle N}_{\rm\scriptscriptstyle N}\right|\xi\right\rangle
\left\langle \xi\left|^{\rm\scriptscriptstyle N}_{\rm\scriptscriptstyle N}\right|\eta\right\rangle
+
\left\langle \eta -\lambda\left|^{\rm\scriptscriptstyle R}_{\rm\scriptscriptstyle R}\right|\xi\right\rangle
\left\langle \xi\left|^{\rm\scriptscriptstyle R}_{\rm\scriptscriptstyle R}\right|\eta\right\rangle
\Big)
=
\int\limits_{-i\infty}^{i\infty}\!\frac{d\rho}{2i}\
I_{1}(\lambda,\rho)\,{\rm e}^{i\pi\rho\eta}
=
8\delta(\lambda).
\end{equation}

Analogous computation gives:
\begin{eqnarray}
\nonumber
&&
\hskip -5mm
\frac{1}{2S^2_{\rm R}(\lambda_\epsilon)}
\int\limits_{-i\infty}^{i\infty}\!\frac{d\tau}{i}\
\int\limits_{-i\infty}^{i\infty}\!\frac{d\xi}{i}\
\nu_{\epsilon}(\xi)
\left[
\left\langle \tau -\lambda_\epsilon\left|^{\rm\scriptscriptstyle R}_{\rm\scriptscriptstyle R}\right|\xi\right\rangle
\left\langle \xi\left|^{\rm\scriptscriptstyle N}_{\rm\scriptscriptstyle N}\right|\tau\right\rangle
-
\left\langle \tau -\lambda_\epsilon\left|^{\rm\scriptscriptstyle N}_{\rm\scriptscriptstyle N}\right|\xi\right\rangle
\left\langle \xi\left|^{\rm\scriptscriptstyle R}_{\rm\scriptscriptstyle R}\right|\tau\right\rangle
\right]
{\rm e}^{-i\pi\rho\tau}
\\[10pt]
\nonumber
& = &
\frac{G_{\rm NS}\left(\frac{\rho-\lambda_\epsilon}{2}+\epsilon\right)}{G_{\rm R}\left(\frac{\rho+\lambda_\epsilon}{2}+\epsilon\right)}
\frac{G_{\rm NS}\left(Q+\frac{\rho-\lambda_\epsilon}{2}-\epsilon\right)}{G_{\rm R}\left(Q+\frac{\rho+\lambda_\epsilon}{2}-\epsilon\right)}
-
\frac{G_{\rm R}\left(b+\frac{\rho-\lambda_\epsilon}{2}\right)}{G_{\rm NS}\left(b+\frac{\rho+\lambda_\epsilon}{2}\right)}
\frac{G_{\rm R}\left(b^{-1}+\frac{\rho-\lambda_\epsilon}{2}\right)}{G_{\rm NS}\left(b^{-1}+\frac{\rho+\lambda_\epsilon}{2}\right)}
\\[10pt]
\nonumber
&+&
\frac{G_{\rm R}\left(\frac{\rho-\lambda_\epsilon}{2}+\epsilon\right)}{G_{\rm NS}\left(\frac{\rho+\lambda_\epsilon}{2}+\epsilon\right)}
\frac{G_{\rm R}\left(Q+\frac{\rho-\lambda_\epsilon}{2}-\epsilon\right)}{G_{\rm NS}\left(Q+\frac{\rho+\lambda_\epsilon}{2}-\epsilon\right)}
-
\frac{G_{\rm NS}\left(b+\frac{\rho-\lambda_\epsilon}{2}\right)}{G_{\rm R}\left(b+\frac{\rho+\lambda_\epsilon}{2}\right)}
\frac{G_{\rm NS}\left(b^{-1}+\frac{\rho-\lambda_\epsilon}{2}\right)}{G_{\rm R}\left(b^{-1}+\frac{\rho+\lambda_\epsilon}{2}\right)}.
\end{eqnarray}
Since in this case there are no poles pinching the integration contours (and
$S_{\rm R}(\lambda)$ is regular for $\lambda = i{\mathbb R}$) we get:
\begin{equation}
\label{completeness:2}
\int\limits_{-i\infty}^{i\infty}\!\frac{d\xi}{i}\
\nu(\xi)
\Big(
\left\langle \eta -\lambda\left|^{\rm\scriptscriptstyle R}_{\rm\scriptscriptstyle R}\right|\xi\right\rangle
\left\langle \xi\left|^{\rm\scriptscriptstyle N}_{\rm\scriptscriptstyle N}\right|\eta\right\rangle
-
\left\langle \eta -\lambda\left|^{\rm\scriptscriptstyle N}_{\rm\scriptscriptstyle N}\right|\xi\right\rangle
\left\langle \xi\left|^{\rm\scriptscriptstyle R}_{\rm\scriptscriptstyle R}\right|\eta\right\rangle
\Big)
\;\ = \;\ 0.
\end{equation}

\section{Discussion}
Construction of the fusion matrix presented in this paper can be placed on a more firm ground
by establishing its relation (in the spirit of \cite{Ponsot:1999uf,Ponsot:2000mt}) to the representation theory
of quantum groups. A natural candidate (see \cite{Jimenez:1990nw})  is $U_q\!\left({\rm osp}(2|1)\right):$
$q$-deformed universal enveloping algebra of  osp$(2|1)$
with a deformation parameter $q={\rm e}^{i\pi b^2}$ \cite{Kulish:1988gr}. Indeed,
generalizing the construction of \cite{Ponsot:2000mt} one can define on $V = L^2({\mathbb R})\times L^2({\mathbb R})$
a continuous series of representations of $U_q\!\left({\rm osp}(2|1)\right)$
with the generators given by:
\begin{eqnarray*}
v_{\alpha}^{(+)}
& = &
{\rm e}^{\pi b x}
\left(
\hskip -5pt
\begin{array}{cc}
0 & \hskip -10pt [\delta_x+Q-\alpha]_{\rm\scriptscriptstyle R}
\\ ~
[\delta_x+Q-\alpha]_{\rm\scriptscriptstyle NS} & \hskip -10pt 0
\end{array}
\right)\!,
\;\;
v_{\alpha}^{(-)}
=
{\rm e}^{-\pi b x}
\left(
\hskip -5pt
\begin{array}{cc}
0 & \hskip -10pt [\delta_x+\alpha-Q]_{\rm\scriptscriptstyle R}
\\ ~
[\delta_x+\alpha-Q]_{\rm\scriptscriptstyle NS} & \hskip -10pt 0
\end{array}
\right)\!,
\end{eqnarray*}
and
\[
K_\alpha = {\sf T}_x^{\frac{ib}{2}}\sigma_0 ,
\]
where
\[
 {\sf T}^{a}_x\,f(x) \;\ = \;\ f\left(x+a\right)
\]
and
\begin{eqnarray*}
[\delta_x+a]_{\rm\scriptscriptstyle R}
& = &
\frac{{\rm e}^{\frac{i\pi ba}{2}}{\sf T}^{\frac{ib}{2}}_x - {\rm e}^{-\frac{i\pi ba}{2}}{\sf T}^{-\frac{ib}{2}}_x}
{{\rm e}^{\frac{i\pi b^2}{2}}- {\rm e}^{-\frac{i\pi b^2}{2}}},
\hskip 1cm
[\delta_x+a]_{\rm\scriptscriptstyle NS}
=
\frac{{\rm e}^{\frac{i\pi ba}{2}}{\sf T}^{\frac{ib}{2}}_x + {\rm e}^{-\frac{i\pi ba}{2}}{\sf T}^{-\frac{ib}{2}}_x}
{{\rm e}^{\frac{i\pi b^2}{2}}+ {\rm e}^{-\frac{i\pi b^2}{2}}}.
\end{eqnarray*}
This representation possesses many virtues analogous to those of the representation of $U_q\!\left({\rm sl}(2,{\mathbb R})\right)$
studied in \cite{Ponsot:2000mt}, which proved to be crucial in relating $U_q\!\left({\rm sl}(2,{\mathbb R})\right)$ to the Liouville theory.
For instance, replacing in $v_{\alpha}^{(\pm)}$ and $K_\alpha$ the parameter
$b$ with $b^{-1},$ we obtain a continuous family of representations (on the same space $V$) of generators of  a ``dual'' quantum supergroup
$U_{\tilde q}\!\left({\rm osp}(2|1)\right)$ with the deformation parameter
\(
\tilde q = {\rm e}^{i\pi b^{-2}}.
\)
Since
\begin{eqnarray*}
{\sf T}^{ib}_{\omega}\,
\frac{S_{\rm NS}(\alpha-i\omega)}{S_{\rm NS}(\bar\alpha-i\omega)}
& = &
\frac{[\alpha-i\omega]_{\rm NS}}{[\bar\alpha-i\omega]_{\rm NS}}\,
\frac{S_{\rm R}(\alpha-i\omega)}{S_{\rm R}(\bar\alpha-i\omega)}\,
{\sf T}^{ib}_{\omega},
\\[4pt]
{\sf T}^{ib}_{\omega}\,
\frac{S_{\rm R}(\alpha-i\omega)}{S_{\rm R}(\bar\alpha-i\omega)}
& = &
\frac{[\alpha-i\omega]_{\rm R}}{[\bar\alpha-i\omega]_{\rm R}}\,
\frac{S_{\rm NS}(\alpha-i\omega)}{S_{\rm NS}(\bar\alpha-i\omega)}\,
{\sf T}^{ib}_{\omega},
\end{eqnarray*}
where
\[
[a]_{\rm\scriptscriptstyle R}
=
\frac{\sin\frac{\pi b a}{2}}{\sin\frac{\pi b^2}{2}},\
\hskip 5mm
[a]_{\rm\scriptscriptstyle NS}
=
\frac{\cos\frac{\pi b a}{2}}{\cos\frac{\pi b^2}{2}},
\]
it is easy to see that for a unitary matrix
\[
\widetilde {\cal I}_{\alpha}
\; = \;
\left(
\begin{array}{cc}
\frac{S_{\rm NS}(\alpha-i\omega)}{S_{\rm NS}(\bar\alpha-i\omega)} & \hskip -5pt 0
\\
0 & \hskip -5pt\frac{S_{\rm R}(\alpha-i\omega)}{S_{\rm R}(\bar\alpha-i\omega)}
\end{array}
\right)
\]
and $\widetilde {\cal O}_\alpha \; = \;\widetilde v_{\alpha}^{(\pm)}\hskip -8pt,\,\ \widetilde K_\alpha$
being Fourier-transformed generators $v_{\alpha}^{(\pm)}\hskip -8pt,\,\ K_\alpha,$
we have:
\[
\widetilde {\cal O}_{Q-\alpha}\,\widetilde {\cal I}_{\alpha} = \widetilde {\cal I}_{\alpha}\,\widetilde {\cal O}_{\alpha},
\]
what proves  equivalence of  representations ${\cal O}_{\alpha}$ and ${\cal O}_{Q-\alpha}.$
Moreover, it turns out to be possible to express a Clebsch-Gordan coefficients for this representation
through a ratios of special functions $S_{\rm NS,R}$ and to relate
the matrix $\mathbf F$ to the Racah-Wigner coefficients (the main technical tools
for this construction are provided by the formulae from Section \ref{special:functions}).
This results will
be reported elsewhere \cite{Hadasz:in:progress}.

Let us conclude with several remarks.

Results from the quantum Liouville theory have a number of applications,
to name only quantization of Teichm\"uller space of Riemann surfaces \cite{Teschner:2003em} and
relation between Liouville theory and the $H_3^+\ $ WZNW model
\cite{Teschner:2001gi,Ponsot:2002cp,Ribault:2005wp,Hosomichi:2006pz,Hikida:2007tq}.
Extension of these results with the help of the results of the present paper
seem to be both  possible and interesting.

The fusion matrix of conformal blocks is related (through the ``renaming'' of variables)
to the three point correlation function of the boundary operators in the Liouville theory
\cite{Teschner:2000md,Ponsot:2001ng}. Once the fusion matrix of the NS blocks is known,
it seems not to be difficult to generalize this link and calculate the (so far unknown) three point function
for the boundary operators in he NS sector of the supersymmetric Liouville theory.

Last but not least, it is plausible that the result of the present paper will allow
to better understand some general properties of the $N=1$ super-conformal Ramond blocks
\cite{Zamolodchikov:1988nm}.

\section*{Acknowledgements}

\noindent
L.H.\ would like to thank the Alexander von Humboldt Foundation for the support during
his stay at  Bonn, R.\ Flume for numerous discussions and encouragement and the faculty
of the Physics Institute if the Bonn University (especially V.\ Rittenberg) for
providing a highly stimulating scientific environment and for a warm hospitality.

\section*{Appendix A. Integral formulae for the funtions $S_{\rm NS,R}(x).$}
\renewcommand{\theequation}{A.\arabic{equation}}
In this Appendix we have collected the integral formulae satisfied by the rations
of two of the special functions $S_{\rm NS,R}(x).$ They can be derived from (\ref{SIRS}) and read:
\begin{eqnarray}
\nonumber
\int\limits_{-i\infty}^{i\infty}\!\frac{d\tau}{i}\
{\rm e}^{\frac{i\pi}{2}\tau\beta}
\frac{S_{\rm R}(\tau + \alpha)}{S_{\rm NS}(\tau + Q)}
& = &
S_{\rm R}(\alpha)
\left[
\frac{G_{\rm NS}\left(\frac{Q+\beta-\alpha}{2}\right)}{G_{\rm R}\left(\frac{Q+\beta+\alpha}{2}\right)}
+
\frac{G_{\rm R}\left(\frac{Q+\beta-\alpha}{2}\right)}{G_{\rm NS}\left(\frac{Q+\beta+\alpha}{2}\right)}
\right],
\\[-4pt]
\label{SNR:1}
\\[-4pt]
\nonumber
\int\limits_{-i\infty}^{i\infty}\!\frac{d\tau}{i}\
{\rm e}^{\frac{i\pi}{2}\tau\beta}
\frac{S_{\rm NS}(\tau + \alpha)}{S_{\rm R}(\tau + Q)}
& = &
-i
S_{\rm R}(\alpha)
\left[
\frac{G_{\rm NS}\left(\frac{Q+\beta-\alpha}{2}\right)}{G_{\rm R}\left(\frac{Q+\beta+\alpha}{2}\right)}
-
\frac{G_{\rm R}\left(\frac{Q+\beta-\alpha}{2}\right)}{G_{\rm NS}\left(\frac{Q+\beta+\alpha}{2}\right)}
\right],
\end{eqnarray}
and
\begin{eqnarray}
\nonumber
\int\limits_{-i\infty}^{i\infty}\!\frac{d\tau}{i}\
{\rm e}^{\frac{i\pi}{2}\tau\beta}
\frac{S_{\rm NS}(\tau + \alpha)}{S_{\rm NS}(\tau + Q)}
& = &
S_{\rm NS}(\alpha)
\left[
\frac{G_{\rm NS}\left(\frac{Q+\beta-\alpha}{2}\right)}{G_{\rm N}\left(\frac{Q+\beta+\alpha}{2}\right)}
+
\frac{G_{\rm R}\left(\frac{Q+\beta-\alpha}{2}\right)}{G_{\rm R}\left(\frac{Q+\beta+\alpha}{2}\right)}
\right],
\\[-4pt]
\label{SNR:2}
\\[-4pt]
\nonumber
\int\limits_{-i\infty}^{i\infty}\!\frac{d\tau}{i}\
{\rm e}^{\frac{i\pi}{2}\tau\beta}
\frac{S_{\rm R}(\tau + \alpha)}{S_{\rm R}(\tau + Q)}
& = &
S_{\rm NS}(\alpha)
\left[
\frac{G_{\rm NS}\left(\frac{Q+\beta-\alpha}{2}\right)}{G_{\rm NS}\left(\frac{Q+\beta+\alpha}{2}\right)}
-
\frac{G_{\rm R}\left(\frac{Q+\beta-\alpha}{2}\right)}{G_{\rm R}\left(\frac{Q+\beta+\alpha}{2}\right)}
\right].
\end{eqnarray}
We shall also need expression for the complex conjugations of the l.h.s. They are of the form:
\begin{eqnarray}
\nonumber
\int\limits_{-i\infty}^{i\infty}\!\frac{d\tau}{i}\
{\rm e}^{-\frac{i\pi}{2}\tau\beta}
\frac{S_{\rm R}(\tau )}{S_{\rm NS}(\tau- \alpha + Q)}
& = &
-i\,S_{\rm R}(\alpha)
\left[
\frac{G_{\rm NS}\left(\frac{Q+\beta-\alpha}{2}\right)}{G_{\rm R}\left(\frac{Q+\beta+\alpha}{2}\right)}
-
\frac{G_{\rm R}\left(\frac{Q+\beta-\alpha}{2}\right)}{G_{\rm NS}\left(\frac{Q+\beta+\alpha}{2}\right)}
\right],
\\[-4pt]
\label{SNR:3}
\\[-4pt]
\nonumber
\int\limits_{-i\infty}^{i\infty}\!\frac{d\tau}{i}\
{\rm e}^{-\frac{i\pi}{2}\tau\beta}
\frac{S_{\rm NS}(\tau )}{S_{\rm R}(\tau- \alpha + Q)}
& = &
S_{\rm R}(\alpha)
\left[
\frac{G_{\rm NS}\left(\frac{Q+\beta-\alpha}{2}\right)}{G_{\rm R}\left(\frac{Q+\beta+\alpha}{2}\right)}
+
\frac{G_{\rm R}\left(\frac{Q+\beta-\alpha}{2}\right)}{G_{\rm NS}\left(\frac{Q+\beta+\alpha}{2}\right)}
\right],
\end{eqnarray}
and
\begin{eqnarray}
\nonumber
\int\limits_{-i\infty}^{i\infty}\!\frac{d\tau}{i}\
{\rm e}^{-\frac{i\pi}{2}\tau\beta}
\frac{S_{\rm NS}(\tau )}{S_{\rm NS}(\tau- \alpha + Q)}
& = &
S_{\rm NS}(\alpha)
\left[
\frac{G_{\rm NS}\left(\frac{Q+\beta-\alpha}{2}\right)}{G_{\rm NS}\left(\frac{Q+\beta+\alpha}{2}\right)}
+
\frac{G_{\rm R}\left(\frac{Q+\beta-\alpha}{2}\right)}{G_{\rm R}\left(\frac{Q+\beta+\alpha}{2}\right)}
\right],
\\[-4pt]
\label{SNR:4}
\\[-4pt]
\nonumber
\int\limits_{-i\infty}^{i\infty}\!\frac{d\tau}{i}\
{\rm e}^{-\frac{i\pi}{2}\tau\beta}
\frac{S_{\rm R}(\tau )}{S_{\rm R}(\tau- \alpha + Q)}
& = &
S_{\rm NS}(\alpha)
\left[
\frac{G_{\rm NS}\left(\frac{Q+\beta-\alpha}{2}\right)}{G_{\rm NS}\left(\frac{Q+\beta+\alpha}{2}\right)}
-
\frac{G_{\rm R}\left(\frac{Q+\beta-\alpha}{2}\right)}{G_{\rm R}\left(\frac{Q+\beta+\alpha}{2}\right)}
\right].
\end{eqnarray}

\end{document}